\documentclass[fleqn,10pt]{wlscirep}
\usepackage[utf8]{inputenc}
\usepackage[T1]{fontenc}
\usepackage{bm}
\usepackage{lscape}
\usepackage{longtable}
\usepackage{array,makecell,multirow,threeparttable}
\usepackage{rotating}
\usepackage{graphicx}

\title{A Systematic Review on Foundation Models for Electrocardiogram Analysis: Initial Strides and Expansive Horizons}

\author[1,2]{Yu Han}
\author[1,3]{Vittorio Murino}
\author[4]{Xiaofeng Liu}
\author[5]{Xiang Zhang}
\author[6,*]{Cheng Ding}
\affil[1]{AI for Good (AIGO), Istituto Italiano di Tecnologia, Genova, Italy}
\affil[2]{University of Genova, Genova, Italy}
\affil[3]{University of Verona, Verona, Italy}
\affil[4]{Yale University, Department of Biomedical Informatics and Data Science, Connecticut, USA}
\affil[5]{ University of North Carolina at Charlotte, Department of Computer Science, Charlotte, USA}
\affil[6]{Nanjing University of Aeronautics and Astronautics, College of Artificial Intelligence, Nanjing, China}
\affil[*]{Corresponding author e-mail: chengding@nuaa.edu.cn}

\begin{abstract}
Electrocardiogram (ECG) is widely used in healthcare applications, such as arrhythmia detection and sleep monitoring, making accurate ECG analysis critically essential. Traditional deep learning models for ECG are task-specific, with limited generalization and narrow functionality. Foundation models (FMs), or large pre-training models, have recently advanced representation learning, enabling strong performance across diverse tasks and motivating their adoption for ECG analysis. Here, we present the first comprehensive review dedicated to ECG foundation models (ECG-FMs). We map the current landscape of architectures, pre-training paradigms, and adaptation strategies, and critically examine their strengths, limitations, and clinical potential. By consolidating this emerging field, we aim to accelerate the development of robust, generalizable ECG-FMs and chart future directions for their integration into healthcare practice.
\end{abstract}
\begin{document}

\flushbottom
\maketitle

\thispagestyle{empty}

\section*{Introduction}

Foundation models (FMs), a new generation of AI models trained on massive and diverse datasets, have advanced rapidly across multiple domains, including language understanding~\cite{kirchenbauer2023watermark, morin2021artificial}, computer vision (CV)~\cite{vemulapalliknowledge}, and multimodal tasks~\cite{fei2022towards, xu2023mplug}. Notable examples include OpenAI's GPT-4~\cite{achiam2023gpt}, a large language model that demonstrates state-of-the-art performance in natural language processing (NLP) tasks such as text generation and summarization, and Meta's LLaMA series~\cite{touvron2023llama, dubey2024llama}, which extends the frontier of open-source large-scale modeling by offering high-performance models trained on diverse data. These models are designed to generalize across tasks, highlighting their flexibility, adaptability, and scalability across a wide range of applications~\cite{guo2024multi, melnyk2023reprogramming}.

In recent years, FMs have attracted growing interest in the medical domain, including electrocardiogram (ECG) analysis~\cite{moor2023foundation, noseworthy2022artificial}. Conventional ECG models are typically designed for narrow, task-specific objectives, relying heavily on labeled datasets and constrained by limited generalizability. By contrast, ECG foundation models (ECG-FMs) offer a more flexible and broadly applicable framework with several notable advantages~\cite{scholte2024scoping}. First, ECG-FMs support a wide range of tasks, from general cardiac disease diagnosis, such as arrhythmia detection~\cite{liu2024learning} and heart rate variability (HRV) analysis~\cite{georgieva2024examination}, to grounded ECG understanding, including predicting heart failure~\cite{raghu2022contrastive} and stratifying patient risk profiles~\cite{wornow2023shaky}. Through pre-training on large and heterogeneous ECG datasets, these models learn transferable representations that exhibit robustness across diverse patient populations, thereby enhancing their utility in real-world clinical environments characterized by substantial variability in data distributions~\cite{qi2022cybertwin}. Second, ECG-FMs enable multimodal integration by combining ECG signals with complementary data sources such as photoplethysmography (PPG)\cite{lan2023performer} or electronic health records (EHRs)\cite{chung2023text}. This integration facilitates a more holistic assessment of patient health and provides richer diagnostic insights~\cite{acosta2022multimodal}.

Despite these advances, prior literature lacks a systematic methodological survey of ECG-FMs. To address this gap, this article provides a comprehensive review that analyzes ECG-FMs from three perspectives: \textit{architecture design}, \textit{pre-training paradigms}, and \textit{adaptation techniques}. This analysis seeks to clarify the mechanisms underpinning the effectiveness of ECG-FMs and to highlight their potential in ECG analysis.

Unlike earlier surveys that focus on general-purpose FMs~\cite{awais2025foundation} or broadly cover medical foundation models~\cite{moor2023foundation, he2024foundation, khan2025comprehensive}, our work concentrates specifically on ECG analysis. Following the categorization in the survey~\cite{liang2024foundation}, we devide ECG-FMs into \textit{Transformer-based models} and \textit{non-Transformer-based models}. Notably, while more than one third of existing studies employ relatively simple architectures such as convolutional neural networks (CNNs) or ResNets, rather than large-scale models like BERTs or LLaMAs, these models still hold significant potential for expansion. With larger datasets and deeper architectures, they can evolve into robust FM frameworks. In this article, we therefore consider both large-scale and simple deep learning models as part of the ECG-FM landscape, underscoring their relevance and versatility. To inspire further innovation, we summarize the development roadmap of current ECG-FMs in Figure 1. Our contributions can be summarized as follows:

\begin{itemize}
	\item \textbf{Comprehensive and up-to-date review.} We provide a thorough and current review of ECG-FMs, including datasets, benchmarks, and a broad spectrum of tasks for ECG analysis.
	\item \textbf{Novel methodology-centric taxonomy.} We introduce a novel taxonomy that examines ECG-FMs from a methodological perspective, with a detailed discussion of architectural design and pipeline workflows.
	\item \textbf{Current challenges and future research opportunities.} We identify key challenges in ECG-FM research from both methodological and clinical viewpoints and outline promising directions for future work, encouraging further exploration of foundation models for ECG analysis.
\end{itemize} 

The rest of the article is organized as follows. Section 2 outlines the literature search strategy. Section 3 introduces the background of FMs. Section 4 reviews the architectures of ECG-FMs, and Section 5 introduces their general development pipeline. Section 6 discusses existing challenges and highlights promising research directions. Finally, Section 7 concludes the article.

\begin{figure}[ht]
\centering
\includegraphics[width=1\linewidth]{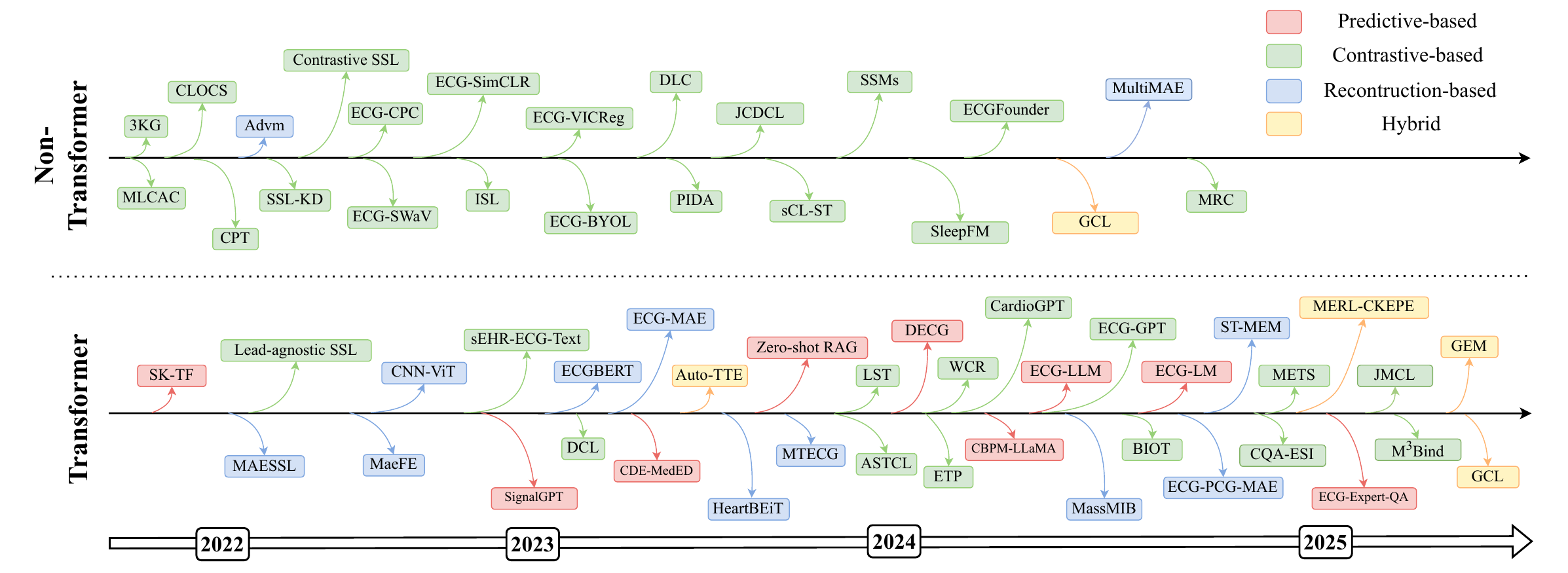}
\caption{\textbf{Roadmaps of representative ECG-FMs.}}
\label{fig:roadmaps_representative_ECG-FMs}
\end{figure}

\section*{Literature search strategy}

We conduct a systematic review of foundation models for ECG analysis following the Preferred Reporting Items for Systematic Reviews and Meta-Analyses (PRISMA) guidelines~\cite{page2021prisma}. Our review focused on English articles published in recent five years, sourced from databases such as PubMed, arXiv, Google Scholar, and publisher platforms like Elsevier, ACM, and IEEE Xplore. To refine our search, we used the following search query: (``foundation model" OR ``large pre-training model" OR ``deep learning" OR ``self-supervised learning") AND (``ECG" OR ``electrocardiogram" OR ``ECG analysis"). Our search yielded a total of 3330 studies, and 66 were retrieved for full-text review after duplicates removal and title and abstract screening. In total, 60 were selected for inclusion in the systematic review, as shown in Figure 2. Studies offering differing perspectives or contradictory evidence were carefully selected to ensure a balanced and comprehensive view of ECG-FMs. The summary of the included studies are shown in Table 1.

\begin{figure}[ht]
\centering
\includegraphics[width=.7\linewidth]{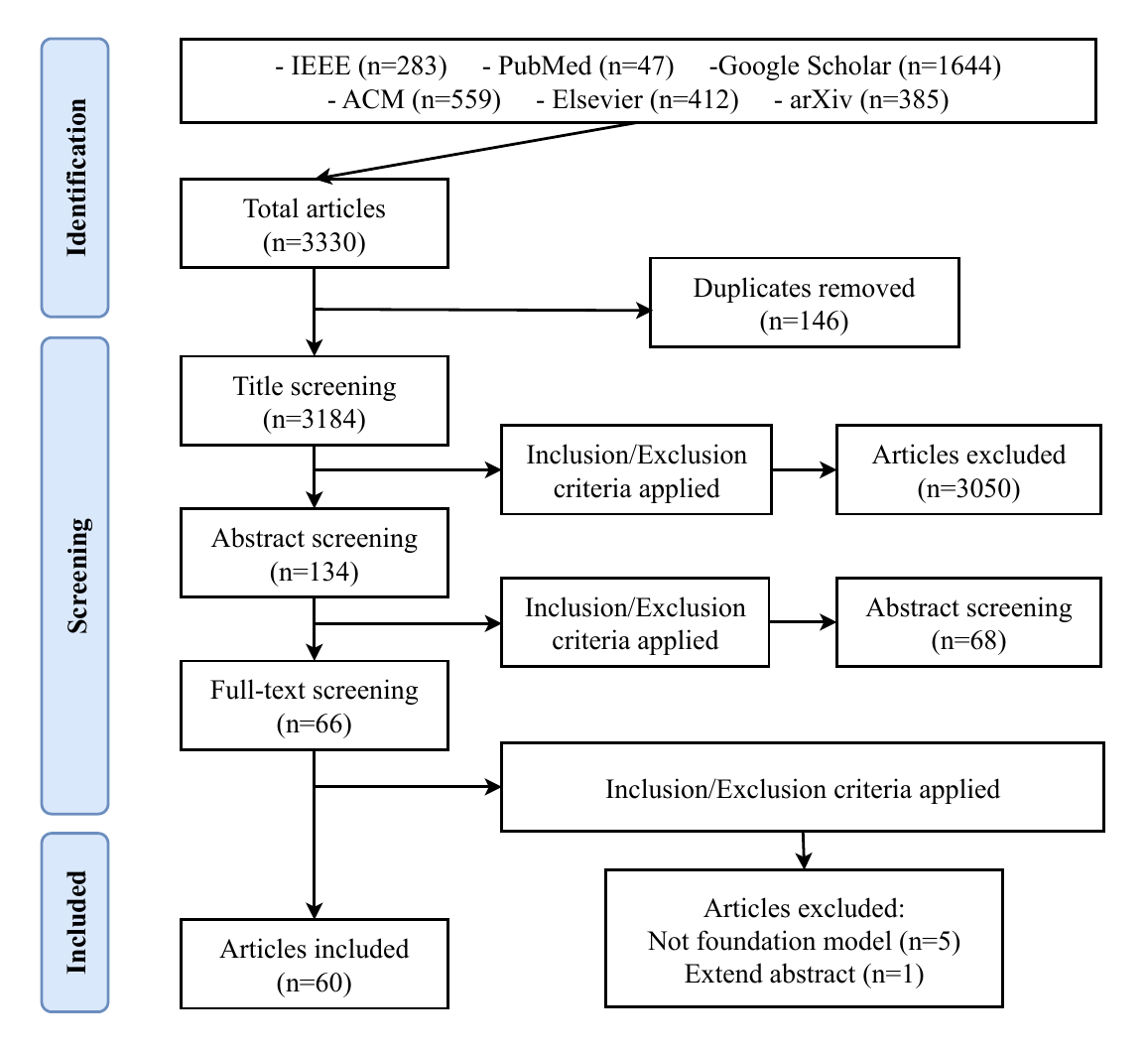}
\caption{\textbf{PRISMA diagram of the literature review process.} We retrieved 60 publications among the 3330 papers that are collected from the six academic platforms.}
\label{fig:PRISMA diagram}
\end{figure}

\section*{Background of foundation models}

FMs are large-scale deep-learning models trained on diverse and extensive datasets in a self-supervised or semi-supervised manner~\cite{bommasani2021opportunities}. FMs generally follow the pre-training and fine-tuning paradigm, which enables the model to learn a broad range of knowledge and patterns and to leverage such prior information in various downstream tasks and applications. FMs only require considerably small-scale custom data in the fine-tuning phase, showcasing remarkable adaptability and efficiency~\cite{wei2022emergent}. FMs have been adopted successfully across domains such as CV, NLP, and healthcare. As the data or model size massively scales up and the available data modality dramatically increases, FMs have evolved from the initial pre-trained models like BERT to contemporary language foundation models (LFMs), vision foundation models (VFMs), vision-language foundation models (VLFMs), and time-series foundation models (TSFMs).

\subsection*{Language foundation models}

In the sphere of NLP, large language models (LLMs) have fundamentally reshaped our understanding of syntactic and semantic information extracted from input text. Autoregressive-based LLMs, exemplified by powerful models like GPT-4 ~\cite{achiam2023gpt}, Qwen2.5~\cite{bai2025qwen2}, and DeepSeek-v3~\cite{liu2024deepseek},  have redefined text generation and question-answering capabilities by predicting subsequent words in a sequence and producing contextually coherent outputs. Additionally, the emergence of the groundbreaking LLM, LLaMA3~\cite{dubey2024llama}, represents a significant advancement, striking a balance between high performance, efficiency, and accessibility through innovative encoding tokenization and memory-efficient attention techniques.
On the other hand, autoencoder-based LLMs, such as BERT~\cite{devlin2018bert} and its remarkable variants like T5~\cite{raffel2020exploring}, RoBERTa~\cite{liu2019roberta}, DistilBERT~\cite{sanh2019distilbert}, and ELECTRA~\cite{clark2020electra}, have redefined the analysis of contextual relationships within text by leveraging bidirectional context consideration. This makes them an indispensable asset for a wide range of NLP tasks, from sentiment analysis to machine translation.

\subsection*{Vision foundation models} 

Inspired by the success of FMs in NLP, practitioners have delved into the realm of vision FMs, encompassing VFMs and VLMs. VFMs, pre-trained on extensive image datasets, excel in comprehending image content and extracting rich semantic information. With the escalating data scale and the growing complexity of deep neural networks, VFMs have evolved from conventional CNN-based models, such as VGG~\cite{simonyan2014very} and EfficientNet~\cite{tan2019efficientnet}, to cutting-edge Transformer-based architectures, including ViT~\cite{dosovitskiy2020image}, Swin Transformer~\cite{liu2021swin}, Linformer~\cite{wang2020linformer}, and DeiT~\cite{touvron2021training}. Moreover, many VFMs employ SSL techniques to acquire generalizable visual representations during pre-training. A prime example is MAE~\cite{he2022masked}, which focuses on the correlation between visible patches and masked regions, enabling it to handle intricate visual structures and capture high-level image semantics.

\subsection*{Vision-language foundation models} 

In addition to VFMs, considerable research efforts have also been channeled into the development of VLMs. VLMs typically incorporate a combined encoder-decoder architecture, leveraging an LLM as the core element, alongside a visual encoder used to transform images into a format compatible with the LLM. For instance, CLIP~\cite{radford2021learning} utilizes contrastive learning methods to create fused representations for images and texts, demonstrating promising capabilities in open-vocabulary image classification. Similarly, SAM~\cite{kirillov2023segment} stands out as a VLM that delivers impressive zero-shot image segmentation performance, enabling class-agnostic segmentation based on the image and a visual prompt specifying the targeted segments. These pre-trained VLMs can then be tailored to specific tasks by presenting them with a natural language description of the task and corresponding prompts. Recent studies have showcased the promising zero-shot performance of pre-trained VLMs across various downstream vision tasks, such as image classification, object detection, and semantic segmentation~\cite{ma2024segment}. Additionally, research endeavors have also explored the development of models aligning multiple paired modalities, including image-text, video-text, and video-audio~\cite{guzhov2022audioclip, ghiasi2022scaling}.

\subsection*{Time series foundation models} 

The increasing richness of time-series data and its wide range of applications have thrust TSFMs into the spotlight as a promising direction for time series analysis~\cite{jin2023large, liang2024foundation}. TSFMs aim to harness the power of the FM paradigm to develop generalized models adept at understanding and forecasting time series data across diverse domains. By leveraging large-scale time series datasets, TSFMs hold the promise of attaining superior performance across a spectrum of time series tasks, offering a unified framework that can expedite research and application development in this field. Based on the nature of time series, TSFMs can be classified into general-purpose TSFMs dealing with standard time series and domain-specific TSFMs focusing on particular tasks such as traffic forecasting and clinical diagnosis. General-purpose TSFMs predominantly utilize single or multiple data modalities to craft robust models targeting specific time series tasks, such as Voice2Series~\cite{yang2021voice2series}, TS2Vec~\cite{yue2022ts2vec}, and SimMTM~\cite{dong2024simmtm} for classification; UniTime~\cite{liu2024unitime} for cross-domain prediction; UniTS~\cite{gao2024units} for multi-task learning; and TimesFM~\cite{das2023decoder}, TimeGPT-1~\cite{garza2023timegpt}, TTMs~\cite{ekambaram2024ttms}, and TSMixer~\cite{ekambaram2023tsmixer} for zero-shot forecasting. Domain-specific TSFMs have distinct focuses and architectures tailored to the requirements of the downstream tasks. In the transportation sector, TSFMs typically employ graph structures, such as graph neural networks (GNNs), to capture the spatio-temporal interactions within dynamic traffic systems. For instance, STGCL~\cite{liu2022contrastive} and SPGCL~\cite{li2022mining} explore the integration of contrastive learning into spatio-temporal graph forecasting, while ST-LLM~\cite{liu2024spatial} combines spatio-temporal information with a partially frozen LLM to enhance traffic predictions. Additionally, TSFMs in the medical domain usually place greater emphasis on capturing the hierarchical characteristics of patient data. For example, FORMED~\cite{huang2024repurposing} can easily capture domain knowledge and be adapted to unseen cohorts with minimum training by repurposing a pre-trained TimesFM. COMET~\cite{wang2024contrast} utilizes a multi-granularity contrastive learning framework to exploit data consistencies across different levels (e.g., observation-level and patient-level) in various medical time series, showcasing its potential benefits in complex disease diagnosis.

\section*{Architecture design of ECG-FMs} 

As shown in Figure 3, we investigate the architectures of existing ECG-FMs, focusing on their backbone models and Transformer modes. This section aims to elucidate the underlying mechanisms that shape the capabilities of ECG-FMs and their applications across diverse ECG-based scenarios.

\begin{figure}[th]
\centering
\includegraphics[width=.9\linewidth]{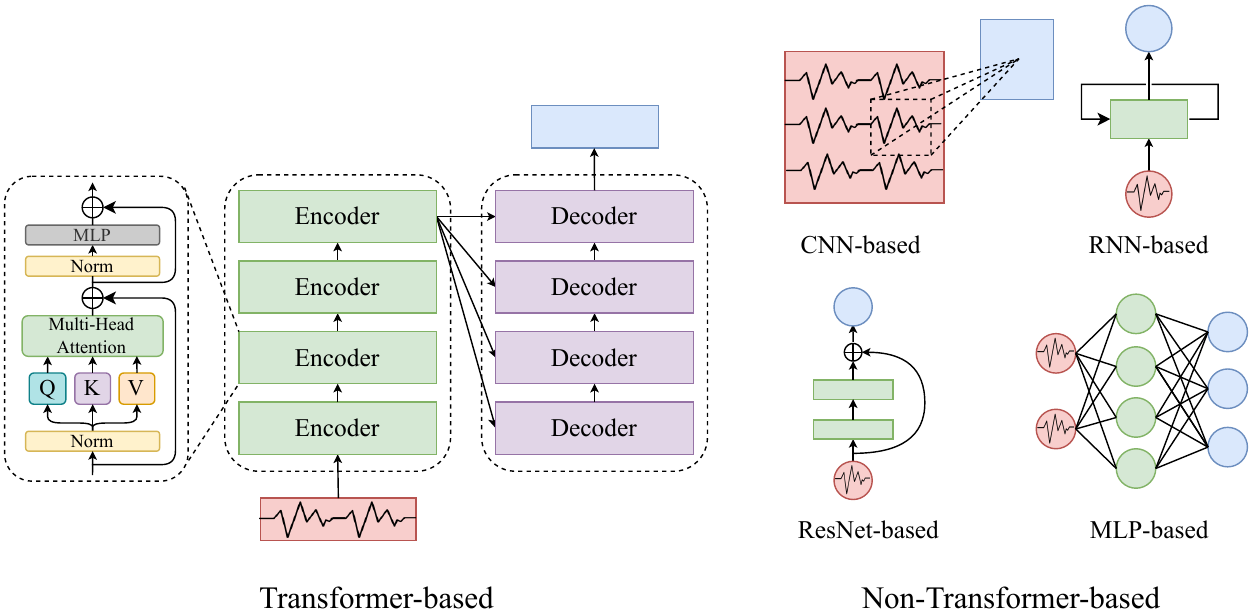}
\caption{\textbf{Architectures of ECG-FMs.}}
\label{fig:architectures_ECG-FMs}
\end{figure}

\subsection*{Transformer-based ECG-FMs}

The architecture of ECG-FMs increasingly builds on the Transformer, originally developed for NLP tasks~\cite{vaswani2017attention}. By relying on attention mechanism rather than convolution or recurrence, Transformers can capture long-range dependencies in sequences and scale effectively across domains, driving their adoption in CV~\cite{kolodiazhnyi2024oneformer3d}, finance~\cite{lin2024fraudgt} and healthcare~\cite{peng2023study}.

While the original Transformer uses an encoder–decoder design, recent studies in ECG analysis have explored three main Transformer modes: \textit{encoder-only}, \textit{decoder-only}, and \textit{encoder–decoder}~\cite{khan2024comprehensive}. Encoder-only models, such as ViTs and BERTs, tokenize the input ECG sequence and use self-attention across all tokens to construct contextualized representations, making them well-suited for tasks that require comprehensive signal understanding, such as cardiac disease detection~\cite{yang2024biot}. Decoder-only models, exemplified by GPTs and LLaMAs, generate outputs autoregressively: each new token is predicted based on previously generated ones. This causal structure enables scalable pre-training on massive unlabeled ECG datasets and supports generative tasks such as ECG synthesis~\cite{chung2023text}, interactive assistants that output text-based interpretations of ECG signals~\cite{liu2023biosignal,yu2024ecg}, and zero-shot inference, i.e., performing tasks without explicit training on them by leveraging representations learned from related tasks~\cite{liu2024zero,yu2023zero}. Hybrid encoder–decoder structures combine both approaches: the encoder first produces a global representation of the ECG signal, which the decoder then uses to autoregressively generate outputs conditioned on this representation. Such designs have been applied to diverse ECG tasks, including automated report generation and sequence-to-sequence prediction~\cite{khunte2024automated,vaid2023foundational,yu2024ecg}. Beyond these hybrids, simpler Transformer variants have also been explored for pre-training~\cite{gaudilliere2021generative,zhou2023masked}. These architectural variations illustrate how different Transformer modes offer complementary strengths and underscore the need for continued exploration in ECG analysis. The following section introduces the commonly used models across the three Transformer modes, highlighting the mainstream backbones of ECG-FMs and illustrating their roles in ECG analysis for clinicians and researchers.

\subsubsection*{Encoder-only ECG-FMs}
The encoder-only ECG-FMs typically include three categories: ViT series, BERT series, and CLIP series.
\begin{itemize}
	\item \textbf{ViT series.} ViT has emerged as a powerful backbone for ECG foundation models because it represents signals as tokens and applies global self-attention to capture both long-range temporal dynamics and local semantic cues~\cite{dosovitskiy2020image}. In ECG tasks, tokens are obtained by segmenting waveforms into time windows or constructing time–lead matrices, with lead and positional embeddings guiding the attention mechanism. Most studies adopt masked autoencoder (MAE) pretraining, where a ViT encoder processes visible tokens and a decoder reconstructs the masked ones, encouraging the encoder to learn transferable spatio-temporal features. Several adaptations highlight this potential. ST-MEM~\cite{na2024guiding} incorporates lead embeddings in a MAE, enabling the ViT encoder to jointly model rhythms and cross-lead relations, improving arrhythmia detection and reduced-lead performance. ECG-MAE~\cite{hu2023spatiotemporal} extended this with a multi-lead ECG ViT, masking across both time and leads to enhance representations for multi-label cardiac diagnosis. Other works have sought to scale this paradigm to larger datasets or to tailor masking strategies for ECG. For example, MaeFE~\cite{zhang2022maefe} applies time, lead, and joint masking with ViT encoders to disentangle temporal and spatial features for tasks such as arrhythmia and myocardial infarction classification. MAESSL~\cite{sawano2022masked} employs ViT encoders of varying scales (ViT-Base, ViT-Large, and ViT-Huge) pretrained with MAE on clinical 12-lead ECGs, then fine-tunes them for left ventricular systolic dysfunction (LVSD) screening across multi-center cohorts. While most approaches rely solely on ViT encoders, some combine them with other feature extractors. CNN-ViT~\cite{yang2022masked} incorporates convolutional modules for local morphological patterns alongside a ViT encoder for long-range context, using masked reconstruction to unify them. LST~\cite{abbaspourazad2023large} further extends ViT-based ECG-FMs to the wearable domain by training on millions of single-lead recordings, confirming their competitiveness with CNN-based encoders.

	\item \textbf{BERT series.} BERT~\cite{devlin2018bert} introduces masked language modeling (MLM) as a pretraining paradigm that learns contextual embeddings by predicting masked tokens from surrounding sequences. Its success in NLP has inspired numerous domain-specific adaptations, such as BEiT~\cite{bao2021beit} for vision, and BioBERT~\cite{lee2020biobert}, ClinicalBERT~\cite{alsentzer2019publicly}, and BioLinkBERT~\cite{yasunaga2022linkbert} for biomedical text. These models share the common principle of leveraging unlabeled corpora through masked prediction, while adjusting tokenization and pretraining corpora to reflect the properties of each domain. When transferred to ECG analysis, BERT-series encoders are generally applied in two ways: (1) as masked representation learners for ECG signals treated as tokenized sequences, or (2) as text encoders in multimodal frameworks that align ECGs with clinical narratives or medical knowledge.

    In vision-inspired applications, BEiT has been repurposed to capture contextual dependencies in ECG images. HeartBEiT~\cite{vaid2023foundational} leverages BEiT-style masked image modeling (MIM) to learn contextual embeddings from tokenized ECG plots, leading to improved diagnostic performance across multiple arrhythmia and structural heart disease tasks. Similarly, ECG-GPT~\cite{khunte2024automated} employs a BEiT encoder pretrained on ImageNet~\cite{deng2009imagenet} for automated diagnostic reporting from ECG images, showing that visual token prediction enables the extraction of rich semantic features relevant to clinical interpretation. These studies illustrate how the MLM-to-MIM transfer can extend naturally from text to vision and then to physiological signal images, with the BEiT providing the attention-based modeling capacity. Beyond images, the adaptation of biomedical BERT variants to multimodal ECG–text learning has become a major trend. ClinicalBERT and BioBERT, pretrained on corpora of clinical notes and biomedical literature respectively, have been widely integrated as textual encoders. ETP~\cite{liu2024etp} aligns ECG signals with textual diagnostic descriptions using ClinicalBERT in a CLIP-style contrastive framework, yielding transferable ECG–text embeddings that generalize across datasets. METS~\cite{li2024frozen} demonstrates that even a frozen ClinicalBERT encoder can provide sufficient language supervision for zero-shot ECG classification, highlighting the value of textual grounding in low-label regimes. These approaches underscore how language-pretrained BERT models enrich ECG foundation models by anchoring them to medical semantics. Meanwhile, some efforts have explored treating ECGs themselves as a type of ``language". For example, ECGBERT~\cite{choi2023ecgbert} directly tokenizes ECG waveforms into discrete segments and applies MLM pretraining, mirroring the original BERT design but over physiological sequences. This framework demonstrates that ECG-specific masked token prediction can uncover latent temporal and morphological patterns, supporting strong transfer to arrhythmia detection and other diagnostic tasks.
	
	\item \textbf{CLIP series.} CLIP~\cite{radford2019language} has demonstrated the power of miltimodal contrastive pretraining by aligning images and text within a shared embedding space. Its dual-encoder design, comprising independent Transformer encoders for vision and language, maximizes similarity between paired samples while minimizing it across negatives, enabling broad zero-shot generalization from natural language prompts. Subsequent adaptations, such as MedCLIP~\cite{wang2022medclip} and BiomedCLIP~\cite{zhang2023biomedclip}, have extended this framework to the medical domain through domain-specific corpora and encoders. Building on these developments, CLIP-like approaches have recently been applied to physiological signals like ECGs, where aligning signals with cardiological text facilitates more interpretable and transferable foundation models.

    Early work explored direct adaptation of CLIP to pair ECGs with textual descriptions. CardioGPT~\cite{khunte2024automated} adopts a CLIP framewrok by coupling an ECG encoder with a GPT-based text decoder, enabling automatic generation of diagnostic reports from raw signals. Similarly, M$^3$Bind~\cite{liu2025multimodal} extends this approach by aligning ECGs with shared text embeddings alongside other medical modalities, demonstrating that even the original CLIP architecture can yield clinically meaningful cross-modal retrieval and interpretation. Building on this foundation, subsequent research developed CLIP-like models tailored to ECG-specific challenges such as label diversity, clinical semantics, and limited dataset size. JMCL~\cite{takahashi2025application} applies large-scale contrastive pretraining on Japanese ECG datasets with over one hundred diagnostic categories, showing that ECG–text alignment enhances classification performance and facilitates transfer across heterogeneous label spaces. ETP~\cite{liu2024etp} further aligns ECG signals with ClinicalBERT-encoded diagnostic reports, producing embeddings that transferred effectively to arrhythmia classification and generalized across datasets METS~\cite{li2024frozen} further freezes the text encoder to reduce complexity while still enabling zero-shot ECG classification guided by textual prompts, highlighting the efficiency of leveraging pretrained language models for ECG–text alignment.
    
\end{itemize}

\subsubsection*{Decoder-only ECG-FMs}
With the rapid progress of LLMs, their scope has extended beyond text-only tasks to encompass multimodal applications. In ECG analysis, both unimodal and multimodal variants (MLLMs) have been investigated, exemplified by models such as Qwen~\cite{yang2025qwen3} and DeepSeek~\cite{liu2024deepseek}. As many of these models are grounded in related architectures and share common design principles, we focus on three representative branches of decoder-only models: GPT series, LLaMA series, and DALL-E series.
\begin{itemize}
	\item \textbf{GPT series.} GPTs are autoregressive Transformers that generate outputs token by token, conditioning each prediction on previously generated tokens through causal self-attention~\cite{radford2018improving}. Unlike encoder-only architectures that learn bidirectional embeddings, GPTs are decoder-only models designed for sequence generation, making them particularly effective for tasks that require producing structured text or reasoning over sequential inputs. In the context of ECG analysis, GPTs are especially attractive because they serve not only as signal encoders but also as generative interpreters, transforming ECG data into clinically meaningful narratives. When appropriately prompted, they can generate waveform descriptions, diagnostic explanations, or step-by-step reasoning, thereby bridging raw signal analysis with natural language communication in ways well aligned with clinical practice.
    
    Recent work has proposed multiple strategies for adapting GPT-style architectures to ECG analysis. One approach leverages frozen GPTs as medical knowledge engines. For instance, by prompting pretrained GPT-4 with ECG-derived features or textualized signal representations, Zero-shot RAG~\cite{yu2023zero} demonstrates that GPTs can generate clinically coherent waveform explanations and support diagnostic reasoning without additional task-specific training. Beyond direct prompting, researchers have also explored GPTs as tools for clinical integration and evaluation. SignalGPT~\cite{liu2023biosignal} incorporates large GPT models into biomedical report-drafting workflows, using their generative capabilities to produce preliminary interpretations that clinicians can refine. ECG-Expert-QA~\cite{wang2025ecg} introduces a benchmark for assessing GPT-based models in heart disease diagnosis, systematically testing their ability to answer cardiology-focused questions and highlighting both their promise and current limitations as diagnostic assistants. Meanwhile, ECG-LM~\cite{yang2025ecg} further extends this direction by training a BioMedGPT directly on ECG data paired with textual annotations, aiming to develop a unified generative model capable of both understanding and producing ECG-related content.
	
	\item \textbf{LLaMA series.} The LLaMA series extends the decoder-only Transformer architecture with efficiency-oriented refinements such as SwiGLU activations~\cite{shazeer2020glu} and rotary positional embeddings (RoPE)~\cite{su2024roformer}. These enhancements enable effective handling of long-context sequences and facilitate parameter-efficient fine-tuning, making LLaMAs particularly well suited for ECG analysis, where extended recordings and computational constraints present persistent challenges. In addition, unlike the GPT series, LLaMA has been released as an open-source framework, providing broad access to large-scale generative modeling. This openness has further stimulated rapid progress in healthcare applications, where transparency and reproducibility are especially critical.

    Recent studies have adapted LLaMA models to diverse ECG applications, including signal restoration, diagnostic reasoning, and wearable health monitoring. For example, ECG-LLM~\cite{liu2024ecg} addresses noisy or incomplete recordings by casting signal restoration as a generative prediction task, in which tokenized ECG inputs are reconstructed by LLaMA3 into physiologically plausible signals suitable for downstream analysis. Similarly, Zero-shot RAG~\cite{yu2023zero} employs LLaMA2 as the reasoning core, where ECG-derived prompts are enriched with retrieved cardiological knowledge and then processed to produce diagnostic outputs without task-specific fine-tuning. Extending to wearable health applications, CBPM-LLaMA~\cite{liu2024large} integrates ECG and PPG signals using LLaMA3, guided by textual supervision, to enable accurate non-invasive blood pressure estimation.
	
	\item \textbf{DALL-E series.} DALL-E~\cite{ramesh2021zero} extends LLMs into text-to-image generation, enabling visual synthesis directly from natural language prompts. Its evolution from the token-based DALL-E-1 to the diffusion-guided DALL-E-2~\cite{ramesh2022hierarchical} and DALL-E-3~\cite{betker2023improvin} has progressively improved visual quality and semantic fidelity, with DALL-E-3 demonstrating particularly strong alignment between text prompts and generated images. While originally developed for creative imagery, these models are increasingly being explored in medicine as tools for producing synthetic clinical illustrations and educational materials.

    In ECG-related research, DALL-E has been investigated primarily for visualization and educational support. For instance, DECG~\cite{zhu2024can} systematically evaluates its capacity to generate realistic ECG waveforms and schematic teaching diagrams from textual prompts, which shows that DALL-E-3 could produce visually convincing ECG-like outputs for instructional purposes, but often fails to capture lead-specific morphology with sufficient precision for clinical diagnosis. CDE-MedED~\cite{amri2023incorporating} demonstrates how DALL-E can support cardiology education by converting textual case descriptions into visual teaching aids, including schematic ECG diagrams and rhythm illustrations, thereby enhancing comprehension and reducing preparation time.
\end{itemize}

\subsubsection*{Encoder-decoder ECG-FMs}
While encoder-only architectures have proven effective for robust ECG representation learning, and decoder-only models have shown strength in generative reasoning and interpretability, many tasks in ECG analysis demand a combination of both capabilities. Encoder–decoder frameworks can naturally provide this balance: the encoder captures contextualized features from the input sequence, while the decoder generates reconstructions, predictions, or textual outputs conditioned on those representations. In ECG-FMs, encoder-decoder designs appear in two main forms: (1) direct use of the original Transformer architecture with both encoder and decoder blocks, and (2) hybrid systems that integrate encoder-only and decoder-only models to leverage complementary strengths.
\begin{itemize}
    \item \textbf{Simple Transformers.} Early studies directly employed the full encoder–decoder Transformer for ECG analysis. For instance, DCL~\cite{lalzary2023dual} utilizes the encoder to extract signal features while the decoder reconstructs masked segments, with dual contrastive objectives enabling representation transfer to emotion recognition and glucose-level estimation. SK-TF~\cite{gaudilliere2021generative} adopts a comparable architecture, pretraining on large ECG datasets before fine-tuning for arrhythmia detection, achieving stronger performance than encoder-only baselines. MTECG~\cite{zhou2023masked} further extends this approach by applying a masking strategy across the encoder–decoder pipeline, where the encoder captures contextual representations and the decoder reconstructs masked portions. Collectively, these studies demonstrate that the simple Transformer remains effective for ECG representation learning by combining bidirectional encoding with reconstruction-driven objectives.

    \item \textbf{Hybrid integrations.} Beyond the classical design, recent research has advanced toward hybrid architectures that integrate encoder-only and decoder-only models, capitalizing on their complementary strengths for ECG analysis. For example, HeartBEiT~\cite{vaid2023foundational} combines a BEiT encoder with a DALL-E decoder. The encoder tokenizes ECG images into patch embeddings using masked image modeling, while the decoder reconstructs the missing patches, compelling the encoder to learn rich spatio-temporal representations. ECG-GPT~\cite{khunte2024automated} links a BEiT encoder with a GPT-2~\cite{radford2019language} decoder, where high-level features extracted by the encoder are passed to the decoder to autoregressively generate textual diagnostic reports. This integration establishes an end-to-end pipeline that translates visual signal inputs into natural language, bridging ECG interpretation with clinical documentation. Building on this framework, other studies have extended the paradigm to multimodal integration. CQA-ESI~\cite{yu2024ecg} first employs a CLIP-like encoder to align ECG signals with biomedical text, and then couples it with a GPT-3.5~\cite{brown2020language} decoder. The encoder produces multimodal embeddings grounded in both signals and text, while the decoder generates potential waveform details and clinical interpretations by querying external medical knowledge bases. Similarly, ECG-Chat~\cite{zhao2024ecg} leverages ECG-CoCa as the encoder to integrate ECG and textual representations into a shared space, with GPT-4o~\cite{achiam2023gpt} serving as the decoder to produce conversational outputs. This design enables interactive diagnostic reasoning and facilitates patient–clinician communication. GEM~\cite{lan2025gem} also employs ECG-CoCa~\cite{zhao2024ecg} for signal encoding, but explores multiple decoder options, including GPT-4o, SFT-LLaVA, and SFT-PULSE~\cite{liu2024teach}. These decoders generate textual reasoning conditioned on multimodal embeddings, thereby supporting feature-grounded analysis, evidence-driven diagnosis, and clinician-style diagnostic workflows.
\end{itemize}

\subsection*{Non-Transformer-based ECG-FMs}

Prior to the widespread acceptance of Transformers, a variety of traditional pre-training methods utilized models such as CNNs, RNNs, MLPs, and their variants as the backbone for pre-training. Each of these models has unique strengths and is known for its effectiveness in obtaining valuable representations.

Both CNNs and MLPs are highly regarded for their ability to effectively model spatial and temporal data. CNN-based architectures have garnered significant attention in self-supervised learning for ECG representation learning, with a notable focus on the utilization of AlexNet~\cite{krizhevsky2012imagenet}, EfficientNet~\cite{tan2019efficientnet}, ResNet~\cite{he2016deep} and VGG~\cite{simonyan2014very} as common backbones of ECG-FMs. These approaches primarily leverage 1D convolutional operations and pooling operations. 1D convolution aids in learning short-term temporal patterns, such as heartbeat mode, while pooling can reduce data dimensionality while retaining essential features like R-peak information. Examples such as JCDCL~\cite{liu2023joint} and 3KG~\cite{gopal20213kg} both utilize a CNN encoder to extract cardiac features, claiming superior efficiency in memory usage and processing speed while delivering competitive performance. Additionally, 1D ResNet is widely used in various pre-training modes, with Advm~\cite{bo2022adversarial, bo2022pretraining}, CPT~\cite{raghu2022contrastive}, GCL~\cite{shi2024universal}, and SSL-KD~\cite{phan2022multimodality} employing different 1D ResNet variants as encoders to extract channel-dependent and periodic ECG patterns. On the other hand, MLP-based models are praised for their lightweight design, offering benefits in terms of reduced computational time and cost. For instance, CPC~\cite{mehari2022self} utilizes a simple MLP to encode input sequences, demonstrating high efficiency in forecasting latent representations.

RNNs and LSTMs are widely acknowledged for their proficiency in modeling long-range temporal data. They excel at capturing dependencies across time steps, making them particularly suitable for analyzing ECG signals where the local information contained within one heartbeat cycle is generally related to earlier parts. The inherent linear complexity of RNN-based architectures provides distinct advantages in efficiently processing long ECG sequences. A notable example is the use of LSTM in CPC~\cite{mehari2022self}. This emerging trend of integrating RNN-based models into the pre-training stage presents an invaluable opportunity for advancing ECG analysis.

\section*{ECG-FM building pipeline}

As shown in Figure 4, we review ECG-FMs from the pipeline perspective, incorporating a wide range of model pre-training and adaptation techniques.

\begin{figure}[th]
	\centering
	\includegraphics[width=1\linewidth]{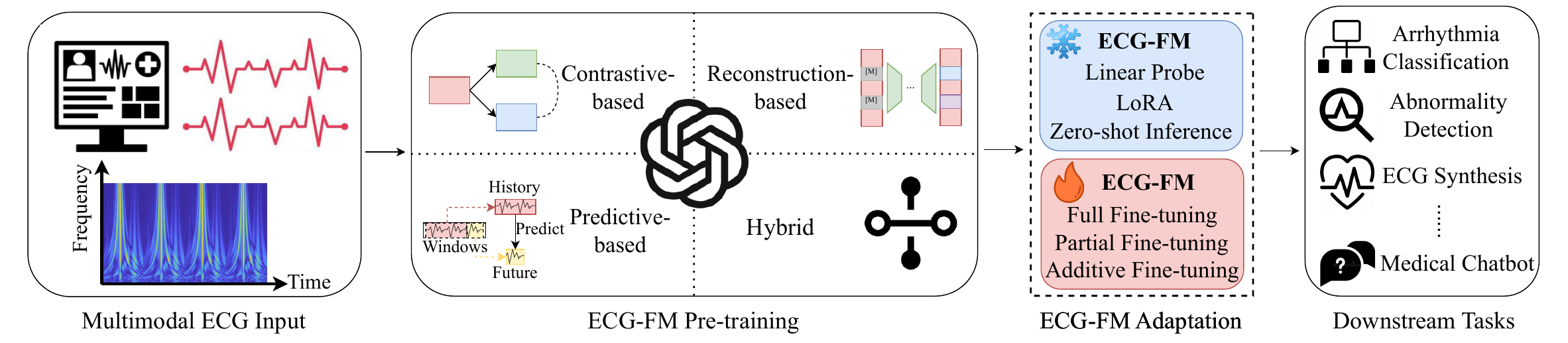}
	\caption{\textbf{Pipeline of building ECG-FMs.}}
	\label{fig:pipeline_ECG-FMs}
\end{figure} 

\subsection*{Pre-training paradigms}

Pre-training represents a crucial first stage in developing ECG-FMs, as it equips the models with the knowledge necessary to adapt swiftly to various downstream tasks and contexts. The diverse modality of pre-training data, encompassing ECG signals, ECG plots, and text-signal/plot pairs, allows for a wide spectrum of pre-training mechanisms in building and deploying ECG-FMs. Notably, ECG plots here refer to visual views of ECG data, where raw signals are treated as 2d images. It is a trick widely used in ECG pre-processing, as it enables the application of advanced vision models to leverage the spatial-temporal structure of ECG data. In this survey, we introduce a new perspective that focuses on learning objectives during pre-training, categorizing existing methods for ECG-FMs into contrastive-based, reconstruction-based, predictive-based, and hybrid techniques.

\subsubsection*{Contrastive-based pre-training}

classify the contrastive-based pre-training strategy as a comprehensive modeling of ECG representations, encompassing \textit{augmentation contrast}, \textit{prediction contrast}, and \textit{multimodal contrast}. Notably, our review adopts a broad perspective by also including contrastive-like models such as BYOL~\cite{grill2020bootstrap} and VICReg~\cite{bardes2021vicreg}, which share the core principle of contrastive pre-training. While these methods are not strictly contrastive in formulation, since they do not rely on explicit negative pairs, they still encourage invariance across different augmented views of the same input. BYOL achieves this by enforcing prediction consistency between an online and a target network, whereas VICReg introduces invariance, variance, and covariance regularization to align augmented representations. Including such methods under a contrastive-like umbrella highlights their conceptual proximity and enables a more unified and systematic discussion of pre-training strategies. As shown in Table 1, augmentation contrast stands as one of the most widely used contrastive frameworks. Most ECG-FMs leverage data augmentation techniques, such as jittering, scaling, and channel masking, to generate different views of an input ECG sample, and then ensure that these augmented views are represented in a similar way in the feature space while pushing representations of different data points far apart. The key lies in constructing and utilizing the self-supervision signals by generating informative positive pairs and filtering out unsuitable negative pairs during augmentation. Due to the periodic nature of the ECG and the significance of the QRS region in containing essential cardiac information, the random heartbeat splitting and joining transformation is also meaningful for ECG augmentation contrast. For instance, sCL-ST~\cite{le2023scl} splits ECG samples into many snippets and randomizes the concatenation of these snippets to produce augmented views, significantly enhancing the robustness of learned representations across different cardiac diseases. Furthermore, augmentation contrast in the frequency domain is also feasible for ECG data. SSL-KD~\cite{phan2022multimodality} leverages fast Fourier transform (FFT) to generate a frequency view for ECG tokens. 

In contrast to data augmentation, prediction contrast takes a unique approach by predicting future information based on the present context. This method aims to maximize the preservation of mutual information between the context and the target, making it highly effective for capturing long-range dependencies in ECG data, such as heart rate variability and changes in waveform shape. SSMs~\cite{mehari2023towards} utilizes the contrastive predictive coding framework to capture semantic information across time for detecting cardiac arrhythmia.

Multimodal contrast generally mines data consistency by considering learning aligned representations across different modalities as the pretext task. The primary goal is to establish a shared representation space where data from various modalities, such as ECG signals and clinical notes, are aligned. During pre-training, the FM is trained to learn aligned representations across various modalities. This involves maximizing the similarity of multimodal views from the same ECG sample while minimizing the similarity of views from different samples. Existing studies typically follow CLIP framework to implement multimodal contrast pre-training. For example, sEHR-ECG-Text~\cite{lalam2023ecg} utilizes a pre-trained BERT model to encode ECG signals, structured EHRs, and unstructured EHRs, resulting in improved performance in ECG classification, zero-shot retrieval, and out-of-distribution detection. Compared to single-modality contrast pre-training methods, multimodal contrast can generate more comprehensive and versatile representations from different modality ECG data, thereby enhancing interpretability and transferability for downstream tasks.

The three branches of contrastive-based pre-training techniques share both similarities and differences. Each branch primarily leverages InfoNCE loss as the contrastive loss function to exploit consistency in ECG samples. However, they differ in terms of the pre-training encoders they utilize. In multi-view augmentation contrast pre-training, typical models such as SimCLR and BYOL are used, while prediction contrast mainly adopts CPC. Additionally, multimodal contrast employs BERT and BEiT variants as the encoders.

\subsubsection*{Reconstruction-based pre-training}

Pre-training in ECG-FM studies generally involves the use of reconstruction techniques, such as \textit{masking} and \textit{autoencoder-based} methods. Masking, also known as masked representation learning, entails retaining a portion of the input and training the model to reconstruct the masked portion. Typically, this process begins by randomly masking a portion of the input ECG data, often by replacing it with a mask token or a zero-value, while keeping the rest of the data visible to the FM. For instance, adversarial masking~\cite{bo2022adversarial, bo2022pretraining} involves using an image-based adversarial masking model to generate mask tokens as augmentations for each given ECG sample during pre-training, effectively reducing distortions of masked ECG signals. MassMIB~\cite{yang2024masked}, employs a random binary masking strategy on segmented patches of both raw ECG signals and the time-frequency spectrograms. BERT along with their variants, are common masking-based pre-training methods in existing ECG-FM studies~\cite{vaid2023foundational}. It typically masks individual tokens, such as ECG timestamps or discrete channels, and leverages bidirectional context and self-attention to predict the masked portion through a Transformer encoder.

Autoencoder-based reconstruction is also essential for developing pre-trained ECG-FMs. Autoencoders are capable of learning a compressed representation of input ECG data by encoding it into a lower-dimensional latent space and then reconstructing the original ECG input from this compressed representation. Models like variational autoencoders (VAEs) and denoising autoencoders (DAEs) have demonstrated outstanding performance in numerous downstream tasks, including ECG synthesis and cardiac disease detection. As an example, Auto-TTE~\cite{chung2023text} harnesses a vector-quantized VAE to compress the raw ECG signals and quantize them into discrete latent representations, thereby enhancing the resilience of synthetic ECG data through the integration of compressed ECG information. Furthermore, masked Autoencoders (MAEs), represent a widely adopted approach in ECG reconstruction, combining a masking strategy with autoencoder architecture~\cite{yang2022masked, mathew2024foundation, hu2023spatiotemporal}. By masking portions of the input ECG signals or plots, MAEs reconstruct the masked regions using an encoder-decoder structure. This technique enables MAE to effectively capture both local details (e.g., individual beats) and global structures (e.g., long-term heartbeat patterns), which is essential for robust representation learning~\cite{sawano2022masked, zhang2022maefe}.

\subsubsection*{Predictive-based pre-training}

In predictive-based pre-training, the FM generates future tokens by conditioning each token on the sequence of previously generated tokens. When applied to ECG analysis, this method typically utilizes Transformer-based models to encode input ECG data, such as heartbeat sequences and image patches, into high-dimensional tokens. These sequential tokens with position information enable FMs to capture temporal and spatial dependencies, as well as long-range relationships across ECG embeddings, like arrhythmias spanning multiple timestamps of recording. Current research commonly employs LLMs like GPTs as pre-trained encoders to obtain robust feature representations for downstream tasks. For instance, zero-shot RAG~\cite{yu2023zero} uses LLaMA2 and GPT-3.5 to convert ECG signals into prompts and integrates expert domain knowledge to guide LLMs in diagnosing arrhythmia and sleep apnea. The specific capability of predictive-based pre-training methods on human instruction comprehension allows ECG-FMs to adapt to question-answering tasks, essentially acting as medical AI bots. SignalGPT~\cite{liu2023biosignal}, for example, offers an interactive interface for users to obtain essential information and fine-tunes ChatGPT by providing feedback. It uses ChatGPT as a comprehensive controller, determining the processing engine according to the signal types and analyzing the description of the given signal, based on pre-training on a massive ECG corpus.

\subsubsection*{Hybrid pre-training}

Several studies have effectively employed hybrid methods that combine contrastive-based, reconstruction-based, and predictive-based techniques in the pre-training process. Hybrid pre-training has been shown to significantly improve the efficiency and robustness of representation learning by integrating the advantages of these methods. First, hybrid pre-training can effectively mitigate the trade-off between sample diversity and the introduction of noise that generally accompanies data augmentation. For example, GCL~\cite{shi2024universal} integrates denoising autoencoder and augmentation contrast, where the decoder reconstructs the augmented views of ECG samples to reduce the noise introduced by data augmentation. Furthermore, hybrid pre-training notably enhances the generalization capability of ECG-FMs by refining the multimodal representations of the samples. A compelling example is CQA-ESI~\cite{yu2024ecg}, which adopts an LLM pipeline to generate detailed textual descriptions of ECG data and leverages contrastive and captioning loss to pre-train ECG encoders for enhanced representations. This approach demonstrates exceptional performance in zero-shot tasks.

\subsection*{Adaptation techniques}

The adaptation paradigm customizes ECG-FM for specific tasks or datasets, thereby enhancing its performance through the utilization of acquired knowledge about cardiac data. Previous methods are classified into four branches: linear probe, fine-tuning, Low-Rank Adaptation (LoRA), and zero-shot inference.

\subsubsection*{Linear probe}

Linear probe is one of the most commonly used strategies for adapting ECG-FMs to downstream tasks. This approach involves training a simple linear classifier, such as logistic regression (LR) or a fully-connected layer (FC), based on the feature representations learned by the pre-trained FM. During the adaptation process, only the weights of the linear classifier are updated through back-propagation, while the remaining FM parameters are kept frozen. The performance of this linear classifier can provide valuable insights into the usefulness of the learned representations for the task at hand. Consequently, it facilitates an assessment of the informational content within the model's learned representations for a specific task or class, without the need for further fine-tuning of the model itself.

\subsubsection*{Fine-tuning}

Fine-tuning serves as a widely utilized strategy for adapting ECG-FMs, with three prevalent approaches: full fine-tuning, partial fine-tuning, and additive fine-tuning. Full fine-tuning involves updating all parameters of the pre-trained model, while partial fine-tuning first freezes the main part of the model and only finetunes the last layer or the last several layers. Additive fine-tuning only selectively updates the parameters of newly introduced layers. Similar to linear probe, most studies employ MLP or ResNet as the additive module. Full fine-tuning necessitates substantial task-specific data and computational resources, making it suitable for tasks significantly different from the initial training tasks, with ample task-specific data available~\cite{shen2021partial}. Conversely, partial fine-tuning and additive fine-tuning offer a more practical option when computational resources are limited or the target task aligns closely with the pre-training tasks~\cite{pittaras2017comparison}.

\subsubsection*{LoRA}

LoRA is an efficient and scalable adaptation approach aimed at reducing the computational burden and memory requirements when adapting large-scale models for various tasks~\cite{hu2021lora}. By employing matrix decomposition techniques, it can efficiently update using two small-scale low-rank matrices. The majority of pre-trained model parameters remain frozen, allowing focused training solely on the additional matrix modules. Renowned for its memory efficiency and adaptability, LoRA has gained popularity in ECG analysis. For instance, Zero-shot RAG~\cite{yu2023zero} employs LoRA to accelerate the adaptation process while preventing overfitting.

\subsubsection*{Zero-shot inference}

Zero-shot inference has emerged as a pivotal adaptation strategy in the realm of ECG-FMs, particularly for LLM-based and VLM-based ECG-FMs. It refers to the capability of the FM to excel at tasks it was not explicitly trained for, without any additional training or fine-tuning on task-specific data. Zero-shot inference typically leverages generalized knowledge to understand and execute new tasks based solely on a description or prompt, without prior examples. For example, METS~\cite{li2024frozen} employs a powerful multimodal contrastive learning framework to capitalize on the similarity between ECG embedding and text embedding, enabling it to accurately compute probabilities for zero-shot ECG classification.

\begin{landscape}
\thispagestyle{fancy} 
{\scriptsize
\renewcommand{\arraystretch}{1.2}
\begin{longtable}{l|m{1.6cm}<{\centering}| m{2.7cm} m{1.8cm} m{1.5cm}  m{1.5cm} m{1.5cm} m{1.5cm} m{1.9cm} m{1.8cm} m{1.5cm} l }  
\caption{Summary of foundation model studies for ECG analysis.}
\\
\hline
\makecell[l]{\bf Pre-Training \\ \bf Category} & \makecell[l]{\bf Pre-Training \\ \bf Subcategory$^{\text{a}}$} &  \textbf{Method} & \textbf{Backbone} & \textbf{Transformer Mode}$^{\text{b}}$ &  \textbf{Adaptation Classifier}$^{\text{c}}$ & \textbf{Adaptation Category}$^{\text{d}}$ & \textbf{Downstream Tasks}$^{\text{e}}$ & \textbf{Pre-Training Datasets} & \textbf{Adaptation Datasets} & \textbf{Data Type} & \makecell[l]{\bf Data \\ \bf  Modality$^{\text{f}}$} \\  
\hline
\endfirsthead
\multicolumn{12}{c}{Table 1: Cont.} \\
\hline
\makecell[l]{\bf Pre-Training \\ \bf Category} & \makecell[l]{\bf Pre-Training \\ \bf Subcategory$^{\text{a}}$} &  \textbf{Method} & \textbf{Backbone} & \textbf{Transformer Mode}$^{\text{b}}$ &  \textbf{Adaptation Classifier}$^{\text{c}}$ & \textbf{Adaptation Category}$^{\text{d}}$ & \textbf{Downstream Tasks}$^{\text{e}}$ & \textbf{Pre-Training Datasets} & \textbf{Adaptation Datasets} & \textbf{Data Type} & \makecell[l]{\bf Data \\ \bf  Modality$^{\text{f}}$} \\  
			\hline
			\endhead 
			\hline
			\endfoot 
			\hline
			\endlastfoot
			\multirow{16}{*}[-40mm]{Contrastive} &  \multirow{16}{*}[-40mm]{AC} &  3KG \cite{gopal20213kg} & CNN1d & N/A & FC & LP & CAC & PhysioNet2020 & PhysioNet2020 & ECG & TS  \\
			\cline{3-12}
			&   &   MLCAC \cite{suh2021learning} & EfficientNet-B3;\newline ResNet-34 & N/A &  FC & LP & CAD & PhysioNet;\newline CinC2021 & PhysioNet;\newline CinC2021 & ECG & TS  \\
			\cline{3-12}
                &   &   \!\footnotemark[1]CLOCS \cite{kiyasseh2021clocs} & CNN1d & N/A & FC;\newline MLP & LP;\newline F-FT & CAC & PhysioNet2020;\newline Chapman & PhysioNet2020;\newline Chapman;\newline PhysioNet2017;\newline Cardiology & ECG & TS \\
                \cline{3-12}
                &   &   CPT \cite{raghu2022contrastive} & ResNet1d & N/A & MLP & F-FT & mPAP;\newline Mortality & MGH & MGH & ECG;\newline Vital signs & TS \\
                \cline{3-12}
			&   &   CPC;\newline SimCLR;\newline BYOL;\newline SWaV \cite{mehari2022self} & xResNet1d-50;\newline LSTM+MLP & N/A &   FC;\newline MLP & LP;\newline F-FT & CAD & CinC2020;\newline Ribeiro;\newline Chapman & PTB-XL & ECG & TS \\
			\cline{3-12}
			&     & Contrastive SSL \cite{rabbani2022contrastive} & CNN1d & N/A & MLP & A-FT & Stress & WESAD;\newline RML & WESAD;\newline RML & ECG & TS  \\
			\cline{3-12}
			&  & \!\footnotemark[2]Lead-agnostic SSL \cite{oh2022lead}  & wav2vec+CMSC & ED &  FC & LP & CAC;\newline Patient & CPSC;\newline CPSC-Extra;\newline PTB-XL;\newline Georgia;\newline NFH;\newline Chapman & CPSC;\newline Georgia & ECG & TS \\
			\cline{3-12}
			&    & ISL \cite{lan2022intra} & ATT-CNN-RNN & N/A &  LR & LP;\newline F-FT & CAC & Chapman;\newline CPSC;\newline PTB-XL & Chapman;\newline CPSC;\newline PTB-XL & ECG & TS  \\
			\cline{3-12}
			&  &   SSL-KD \cite{phan2022multimodality} & SE-ResNet34 & N/A &   MLP & F-FT & CAC & PhysioNet2020 & PhysioNet2020 & ECG & TS+TF  \\
			\cline{3-12}
			&  &  SimCLR;\newline BYOL;\newline VICReg \cite{andersson2023augmentation}  & ResNet-50 & N/A &   ResNet1d-50 & A-FT & CAC & CinC2020;\newline Ribeiro;\newline Chapman & PTB-XL & ECG & TS  \\
			\cline{3-12}
			&  &   SimCLR;\newline BYOL;\newline SWaV \cite{soltanieh2023distribution} & xResNet1d-50 & N/A &   FC & LP & CAC & PTB-XL;\newline Chapman & PTB-XL;\newline Chapman;\newline Ribeiro & ECG & TS  \\
			\cline{3-12}
			&  & DCL \cite{lalzary2023dual} & CNN1d\newline +Transformer & ED &   FC & LP & Emotion;\newline BGL & MAHNOB-HCI;\newline WESAD;\newline CASE;\newline PsPM-FR;\newline  PsPM-RRM1-2;\newline ASCERTAIN & DREAMER;\newline D1NAMO & ECG & TS  \\
			\cline{3-12}
			&   & DLC \cite{liu2023dense} & AlexNet1d;\newline VGGNet1d;\newline ResNet1d & N/A &  FC & LP;\newline F-FT & CAD & NFH;\newline SNPH & PTB-XL;\newline CPSC2018;\newline Chapman & ECG & TS \\
			\cline{3-12}
			&  &   \!\footnotemark[3]PIDA \cite{lai2023practical} & MSDNN+MoCo & N/A &  MSDNN & F-FT & CAD & Cardiocloud W12L-ECG & Cardiocloud W12L-ECG & ECG & TS \\
			\cline{3-12}
			&  &   sCL-ST \cite{le2023scl} & xResNet1d-50 & N/A &   FC & F-FT & CAC & CSPC;\newline INCART;\newline G12EC;\newline PTB & PTB-XL & ECG & TS \\
			\cline{3-12}
			&   &   LST \cite{abbaspourazad2023large} & EfficientNet1d;\newline ResNet1d;\newline ViT1d & EO &  FC & LR & Age;\newline BMI;\newline Gender & AHMS & AHMS & ECG;\newline PPG & TS  \\
			\cline{3-12}
			&  & SimCLR \cite{nath2023towards} & CNN1d & N/A &  FC & LP & Stress & N/A & N/A & ECG & TS  \\
			\cline{3-12}
			&   &   ASTCL \cite{wang2023adversarial} & CNN1d\newline +Transformer & ED &  FC & LP;\newline F-FT & CAC & Chapman;\newline PTB-XL;\newline CODE;\newline CPSC2018;\newline CMI & Chapman;\newline PTB-XL;\newline CODE;\newline CPSC2018;\newline CMI & ECG & TS  \\
			\cline{3-12}
			\multirow{14}{*}[-30mm]{Contrastive}  &  \multirow{5}{*}[-10mm]{AC} &   SleepFM \cite{thapa2024sleepfm} & CNN & N/A &  LR & LP & Sleep stage;\newline Sleep apnea & Multi-modal PSG & Multi-modal PSG & EEG;\newline ECG;\newline Respiratory & TS  \\
			\cline{3-12}
			&   &  MRC \cite{liu2024learning} & CNN & N/A &  FC & LP;\newline F-FT & CAC & NFH & PTB-XL;\newline CPSC2018;\newline Chapman & ECG & TS  \\
			\cline{3-12}
			&   &   \!\footnotemark[4]BIOT \cite{yang2024biot} & Linear\newline Transformer & EO &  MLP & F-FT & CAD & SHHS;\newline PREST;\newline Cardiology;\newline CHB-MIT;\newline IIIC Seizure;\newline TUAB;\newline TUEV;\newline PTB-XL;\newline HAR & CHB-MIT;\newline IIIC Seizure;\newline PTB-XL;\newline HAR & ECG;\newline EEG;\newline HAR & TS+FD  \\
			\cline{3-12}
			&   &   \!\footnotemark[5]WCR \cite{mckeen2024ecg} & wav2vec+CMSC & EO &  FC & LP & LVEF;\newline ACTN;\newline CAD & UHN-ECG;\newline PhysioNet2021;\newline MIMIC-IV-ECG & UHN-ECG & ECG & TS  \\
			\cline{3-12}
			&   &   ECGFounder \cite{li2024electrocardiogram} & RegNet & N/A &   FC & LP & CAC;\newline CAD;\newline Demographics;\newline CED & HEEDB;\newline CODE-test;\newline PTB-XL;\newline PhysioNet2017 & MIMIC-IV-ECG;\newline DeepBeat & ECG;\newline PPG & TS  \\
			\cline{2-12}
			&  \multirow{7}{*}[-15mm]{MC} & sEHR-ECG-Text \cite{lalam2023ecg} & BERT;\newline CNN;\newline GatorTron & EO &  LR;\newline MLP & LP;\newline A-FT & CAC;\newline OOD & Mayo Clinic EHRs & Mayo Clinic EHRs;\newline PhysioNet2020;\newline Chapman & ECG;\newline EHRs & TS+Text  \\
			\cline{3-12}
			&   &    \!\footnotemark[6]JCDCL \cite{liu2023joint} & CNN & N/A &  FC & LP & CAD & NFH & PTB-XL;\newline CPSC2018 & ECG & TS+Plot  \\
			\cline{3-12}
			&  & CardioGPT \cite{fu2024cardiogpt} & CLIP & EO &  FC & LP & CAD & GE\&Philips;\newline NSH & GE\&Philips;\newline NSH & ECG;\newline ECG reports & TF+Text  \\
			\cline{3-12}
			&  & ETP \cite{liu2024etp} & ResNet1d-18+ClinicalBERT & EO &  FC & LP;\newline ZS & CAC;\newline ZSC & PTB-XL;\newline CPSC2018 & PTB-XL;\newline CPSC2018 & ECG;\newline ECG reports & TS+Text  \\
			\cline{3-12}
			&  & ECG-GPT \cite{khunte2024automated} & BEiT+GPT2 & ED &  DistilBERT & P-FT & CAD & YNHHS & UK-Biobank;\newline PTB-XL;\newline Ribeiro & ECG;\newline ECG reports & Text+Plot  \\
			\cline{3-12}
			&  &  METS  \cite{li2024frozen} & ResNet1d-18+ ClinicalBERT & EO &  FC & LP;\newline ZS & ZSC & MIMIC-III;\newline PTB-XL & PTB-XL;\newline MIT-BIH & ECG;\newline ECG reports & TS+Text  \\
			\cline{3-12}
			&  &  CQA-ESI \cite{yu2024ecg} & ConvNext-v2;\newline BioLinkBERT;\newline GPT-3.5 & EO &  FC & LP;\newline F-FT & CAC;\newline Patient;\newline ECG QA & PTB-XL;\newline Chapman;\newline MIMIC-IV-ECG & PTB-XL;\newline ICBEB & ECG;\newline ECG reports & TS+Text  \\
                \cline{3-12}
                &  &  JMCL \cite{takahashi2025application} & ResNet1d+\newline MedLlama3-JP-v2text & EO & FC & ZS & ZSC & UTH;\newline Mitsui & UTH;\newline Mitsui & ECG;\newline ECG reports & TS+Text  \\
			\cline{3-12}
                &  &  M$^3$Bind \cite{liu2025multimodal} & ECG-CLIP & EO & LoRA & ZS & ZSC & MIMIC-CXR;\newline Quilt-1M & MESIDOR;\newline ODIR;\newline CheXpert;\newline RSNA;\newline MIMIC-ECG;\newline PTB-XL;\newline ICBEB;\newline SkinCancer;\newline Camelyon;\newline SICAPv2;\newline SkinTumor;\newline CT-RATE;\newline RAD-ChestCT & ECG;\newline Retina;\newline X-ray;\newline CT;\newline Pathology;\newline Reports & Image+Text  \\
			\cline{2-12}
			& PC & \!\footnotemark[7]SSMs \cite{mehari2023towards} & S4 with FCE & N/A &   MLP & A-FT & CAC & CinC & PTB-XL;\newline Chapman & ECG & TS  \\
\hline
\multirow{12}{*}[-35mm]{Reconstruction}  & \multirow{6}{*}[-10mm]{AE} & MAESSL \cite{sawano2022masked} & CNN;\newline ViT-Large & EO &  MLP & F-FT & LVSD & N/A & N/A & ECG & TS \\[2mm]
\cline{3-12}
&  &   \multirow{2}{*}{MaeFE \cite{zhang2022maefe}} & ViT1d;\newline Sub2D-ViT & \multirow{2}{*}{EO} &  \multirow{2}{*}{FC} & LP;\newline F-FT & \multirow{2}{*}{CAC} & CPSC2018;\newline PTB-XL;\newline Ningbo & PTB-XL;\newline CPSC2018 & ECG & TS+TF  \\
&  &   &   &   &   &  &   &   &   &   &  \\
\cline{3-12}
  &  & CNN-ViT \cite{yang2022masked} & Res2Net+SE\newline +ViT & EO  &  FC & F-FT & CAC & PTB-XL;\newline Chapman;\newline CPSC2018 & PTB-XL;\newline Chapman;\newline CPSC2018 & ECG & TS  \\
			\cline{3-12}
			&  &  ECG-MAE \cite{hu2023spatiotemporal} & MLE-ViT & EO &  FC & LP;\newline F-FT & CAD & CinC2020;\newline Chapman & PTB-XL & ECG & TS  \\
			\cline{3-12}
			&   &  ECG-PCG-MAE \cite{mathew2024foundation} & ViT & EO &  FC & LP & SHDM;\newline AF;\newline LEF & AFib;\newline SHD;\newline Low-EF & AFib;\newline SHD;\newline Low-EF & ECG;\newline PCG & TS+FD  \\
			\cline{2-12}
			&  \multirow{6}{*}[-25mm]{M} &   \!\footnotemark[8]Advm \cite{bo2022adversarial} & ResNet1d-18 & N/A &   FC & LP & CAC;\newline Age & PhysioNet/ CinC2020 & Chapman & ECG & TS  \\
			\cline{3-12}
			&  &   \!\footnotemark[8]Advm \cite{bo2022pretraining} & ResNet1d-18 & N/A &   FC & LR & CAC;\newline Gender & PhysioNet/ CinC2020 & Chapman & ECG & TS  \\
			\cline{3-12}
			&  & \!\footnotemark[9]HeartBEiT \cite{vaid2023foundational} & DALL-E+BEiT & ED &   MLP & F-FT & LVEF;\newline HCM;\newline STEMI & MSHS & MSHS;\newline PTB-XL & ECG & Plot \\
			\cline{3-12}
			&  & MTECG \cite{zhou2023masked} & Transformer & ED &   FC & A-FT & CAC & Fuwai;\newline PTB-XL;\newline PCinC & Fuwai;\newline PTB-XL;\newline PCinC & ECG & Plot  \\
			\cline{3-12}
                &   &   ECGBERT \cite{choi2023ecgbert} & BERT & EO & MLP & F-FT & AF;\newline Sleep apnea;\newline Heartbeat;\newline Patient & MIMIC-III;\newline PTB-XL;\newline Georgia;\newline CPSC-2018 & MIT-BIH & ECG & TS \\
                \cline{3-12}
                &   &   MultiMAE \cite{fang2024promoting} & MAE & N/A & FC & LP & Sleep;\newline Age;\newline Arousal & PhysioNet2018 & PhysioNet2018 & ECG;\newline EEG;\newline EOG;\newline EMG & TS \\
                \cline{3-12}
			&  &  \!\footnotemark[10]MassMIB \cite{yang2024masked} & CNN+ViT & EO &   FC & F-FT & CAD & PTB-XL;\newline CPSC;\newline Chapman & PTB-XL;\newline CPSC;\newline Chapman & ECG & TF  \\
			\cline{3-12}
			&  &  ST-MEM \cite{na2024guiding} & ViT-B & EO &   FC & LP;\newline F-FT & CAC;\newline MIC & Chapman;\newline Ningbo;\newline CODE-15 & PTB-XL;\newline CPSC2018;\newline PhysioNet2017 & ECG & Plot \\
			\hline
			\multirow{3}{*}[-30mm]{Predictive}  & \multirow{3}{*}[-30mm]{AR}  & SK-TF \cite{gaudilliere2021generative} & Transformer & ED &   FC;\newline LR & LP;\newline F-FT & CAC & St-Petersburg;\newline PTB-XL;\newline Georgia;\newline Chapman;\newline Ningbo & CinC2021 & ECG & TS  \\
			\cline{3-12}
			&   & SignalGPT \cite{liu2023biosignal} & ChatGPT & DO &   MLP & P-FT & CAD;\newline ECG QA & CODE-test & CODE-test & ECG;\newline EEG;\newline EMG;\newline EOG & TS  \\
			\cline{3-12}
			&   & Zero-shot RAG \cite{yu2023zero} & LLaMA2;\newline GPT-3.5 & DO &   N/A & LoRA;\newline ZS & CAC;\newline ZSC;\newline Sleep apnea & PTB-XL+;\newline Apnea-ECG & PTB-XL+;\newline Apnea-ECG & ECG;\newline ECG reports & TS+Text  \\
                \cline{3-12}
                &   & CDE-MedED \cite{amri2023incorporating} & ChatGPT;\newline DALL-E & DO & N/A & ZS & Synthesis & N/A & N/A & ECG;\newline X-ray & Image  \\
                \cline{3-12}
                &   & DECG \cite{zhu2024can} & DALL-E & DO & N/A & ZS & Synthesis & N/A & N/A & ECG & ECG plot  \\
                \cline{3-12}
                &   & CBPM-LLaMA \cite{liu2024large} & LLaMA3 & DO & N/A & F-FT & CBPM & N/A & CAS-BP & ECG;\newline PPG & TS  \\
                \cline{3-12}
                &   &   ECG-LM \cite{yang2025ecg} & BioMedGPT-LM;\newline GPT-3.5 & DO & FC & F-FT & ECG QA;\newline CAD;\newline ZSC & N/A & PTB-XL;\newline PTB-XL+;\newline ECG-QA & ECG & TS+Text \\
                \cline{3-12}
                &   &   \!\footnotemark[11]ECG-LLM \cite{liu2024ecg} & LLaMA2;\newline LLaMA3 & DO & MLP & F-FT & Forecasting;\newline Imputation & N/A & PhysioNet2020 & ECG & TS \\
                \cline{3-12}
                &   &   \!\footnotemark[12]ECG-Expert-QA \cite{wang2025ecg} & GPT-4o;\newline DeepSeek-v3;\newline Qwen-2.5;\newline MiniMind-2 & DO & N/A & ZS & ECG QA & N/A & ECG-Expert-QA & ECG;\newline ECG reports & TS+Text \\
			\hline
			\multirow{5}{*}[-12mm]{Hybrid} & AE+AR  & \!\footnotemark[13]Auto-TTE \cite{chung2023text} & VQ-VAE & DO &   N/A & N/A & Synthesis & PTB-XL;\newline Sejong &  PTB-XL;\newline Sejong & ECG;\newline ECG reports & TS+Text  \\
			\cline{2-12}
			& AE+AC  & GCL \cite{shi2024universal} & ResNet1d;\newline VGGNet1d & N/A &   FC & LP;\newline F-FT & CAC;\newline Denoising & SNPH & PTB-XL;\newline Chapman;\newline CPSC2018 & ECG & TS \\
			\cline{2-12}
			& AR+M+MC  & MERL-CKEPE \cite{liu2024zero} & ResNet1d-18+ Med-CPT & DO &   FC & LP;\newline ZS & CAC;\newline MIC;\newline ZSC & MIMIC-ECG & PTB-XL;\newline CPSC2018;\newline Chapman;\newline Ningbo & ECG;\newline ECG reports & TS+Text  \\
                \cline{2-12}
                & \multirow{2}{*}[-2mm]{AR+MC}  & \!\footnotemark[14]GEM \cite{lan2025gem} & ECG-CoCa;\newline GPT-4o;\newline SFT-LLaVA;\newline SFT-PULSE & ED & MLP & LoRA & Grounded\newline understanding & MIMIC-IV-ECG;\newline PULSE;\newline ECG-Bench & MIMIC-IV-ECG;\newline PULSE;\newline ECG-Bench & ECG & TS+Plot+Text \\
                \cline{3-12}
                &   & \!\footnotemark[15]ECG-Chat \cite{zhao2024ecg} & ECG-CoCa;\newline GPT-4o & ED & FC & LoRA & Report\newline generation & N/A & ECG-Instruct & ECG;\newline ECG reports & TS+Text \\
			\hline                  
		\end{longtable}
		\begin{threeparttable}
			\vspace{-8mm}
			\begin{tablenotes}  \tiny   
				\item[a] \textit{AC} Augmentation Contrast, \textit{MC} Multimodal Contrast, \textit{AC} Augmentation Contrast, \textit{AR} Autoregressive-based, \textit{M} Masking, \textit{AE} Autoencoder-based, \textit{PC} Prediction Contrast.
				\item[b] \textit{EO} Encoder-only, \textit{DO} Decoder-only, \textit{ED} Encoder-decoder.
				\item[c] \textit{FC} Fully-connected Layer, \textit{LR} Linear Regression, \textit{MLP} Multilayer Perceptron.
				\item[d] \textit{LP} Linear Probe, \textit{F-FT} Full Fine-tuning, \textit{P-FT} Partial Fine-tuning, \textit{A-FT} Additive Fine-tuning, \textit{ZS} Zero-shot, \textit{LoRA} Low-Rank Adaptation.
				\item[e] \textit{CAC} Cardiac Arrhythmia Classification, \textit{mPAP} Mean Pulmonary Arterial Pressure Detection, \textit{Mortality} Mortality Prediction, \textit{CAD} Cardiac Abnormality Detection, \textit{CED} clinical Event Detection, \textit{MIC} Myocardial Infarction Classification, \textit{Stress} Stress Detection, \textit{BGL} Blood Glucose Level Detection, \textit{Patient} Patient Identification, \textit{ZSC} Zero-shot Classification, \textit{Synthesis} ECG Synthesis, \textit{CBPM} Cuffless Blood Pressure Measurement, \textit{Heartbeat} Heartbeat Classification, \textit{Sleep stage} Sleep Stage Classification, \textit{Sleep apnea} Sleep Apnea Detection, \textit{Arousal} Arousal Identification, \textit{OOD} Out-of-Distribution Detection, \textit{Emotion} Emotion Recognition, \textit{BGL} Blood Glucose Level Monitoring, \textit{Denoising} ECG Denoising, \textit{ACTN} Abnormal Cardiac Troponin Diagnosis, \textit{LVSD} Left Ventricular Systolic Dysfunction Diagnosis, \textit{LVEF} Left Ventricular Ejection Fraction Diagnosis, \textit{HCM} Hypertrophic Cardiomyopathy Diagnosis, \textit{STEMI} ST-elevation Myocardial Infarction Diagnosis, \textit{SHDM} Structural Heart Disease Murmurs Diagnosis, \textit{AF} Atrial Fibrillation Diagnosis, \textit{LEF} Low Ejection Fraction Diagnosis, \textit{Gender} Gender Classification, \textit{Age} Age Classification, \textit{BMI} Body Mass Index Classification, \textit{ECG QA} ECG Question \& Answering.
				\item[f] \textit{TS} Time Series, \textit{Plot} ECG Plot, \textit{TF} Time-Frequency Image, \textit{FD} Frequency Domain.
                \item[1] \url{https://github.com/danikiyasseh/CLOCS}
				\item[2] \url{https://github.com/Jwoo5/fairseq-signals}
				\item[3] \url{https://github.com/SMU-MedicalVision/ECG-Classfication}
				\item[4] \url{https://github.com/ycq091044/BIOT}
				\item[5] \url{https://github.com/bowang-lab/ECG-FM}
				\item[6] \url{https://github.com/Aiwiscal/CDCL-ECG}
				\item[7] \url{https://github.com/tmehari/ssm_ecg}
				\item[8] \url{https://github.com/jessica-bo/advmask_ecg}
				\item[9] \url{https://github.com/akhilvaid/HeartBEiT}
				\item[10] \url{https://github.com/ysxGitHub/MassMIB}
                \item[11] \url{https://github.com/dragonlfy/ECG-LLM}
                \item[12] \url{https://github.com/Zaozzz/ECG-Expert-QA}
                \item[13] \url{https://github.com/TClife/text_to_ecg}
                \item[14] \url{https://github.com/lanxiang1017/GEM}
                \item[15] \url{https://github.com/YubaoZhao/ECG-Chat}
			\end{tablenotes}
		\end{threeparttable}  
	}
\end{landscape}

\section*{Discussion}

Our discussion highlights both challenges and opportunities. In addition to addressing the key barriers hindering the clinical adoption of ECG-FMs, we underscore important findings that advance ECG interpretation and support broader healthcare tasks.

\subsection*{Multimodal inputs help understand ECG data.}

Recent advances in ECG-FMs demonstrate that performance depends not only on model architecture but also on input representation and the richness of available information. While many models directly process raw ECG signals, others exploit transformed representations such as ECG plots, spectrograms, or time–frequency images. These variations underscore the need to carefully consider data format, as it shapes both model design and performance. For example, MTECG~\cite{zhou2023masked} converts ECG segments into image-like patches and applies masked autoencoders to reduce redundancy and enhance wave-shape feature learning, whereas BIOT~\cite{yang2024biot} employs a frequency-domain tokenization strategy using Fourier-based energy vectors to capture spectral dynamics and enable robust cross-dataset pre-training. However, whether raw signals or transformed representations provide a superior foundation for ECG representation learning remains an open question.

Beyond input representation, multimodal integration has emerged as a promising approach to enrich ECG understanding. In addition to ECG, several studies recent studies incorporate related physiological signals (e.g., PPG, EMG) and complementary information (e.g., EHRs, spectrograms). By combining multiple modalities, these frameworks capture a more holistic view of cardiovascular function and patient state. Empirical evidence shows that multimodal models consistently outperform unimodal counterparts across downstream tasks and model architectures. For instance, SSL-KD~\cite{phan2022multimodality} fuses ECG series with spectrograms for arrhythmia classification, boosting accuracy from 48.8\% to 48.9\% and F1-score from 61.4\% to 62.1\% on PhysioNet2020; ECG-PCG-MAE~\cite{mathew2024foundation} achieves an average Area Under the ROC Curve (AUC) gain of 5.5\% for LEF detection compared to ECG-only or PCG-only models; GEM~\cite{lan2025gem} integrates raw ECG, transformed images, and textual instructions to yield a 5.8\% improvement for grounded abnormality detection on PTB-XL and CODE. Overall, these findings suggest that multimodal frameworks not only enhance predictive performance but also strengthen interpretability and alignment with clinical reasoning. Each modality provides unique and complementary cues: for example, spectrograms highlight frequency-domain patterns such as rhythm regularity or noise artifacts that may be overlooked in raw ECG; PCG captures mechanical heart sounds indicative of valve function; PPG reflects peripheral vascular responses and oxygenation dynamics; and EHRs encode semantic knowledge of patient history, comorbidities, and treatments. By unifying these heterogeneous sources, ECG-FMs can learn richer latent representations that jointly capture electrophysiological, hemodynamic, and contextual dimensions of cardiovascular health. This integration not only mitigates sensitivity to noise and artifacts in any single modality, but also aligns model reasoning more closely with real-world clinical practice, where clinicians routinely synthesize heterogeneous evidence from multiple streams to reach informed decisions.

\subsection*{ECG-FMs still face barriers to clinical workflow integration.}

Despite the strong performance across benchmark tasks, most ECG-FMs are still evaluated in controlled, offline settings and have not yet been embedded into routine clinical workflows. However, real-world deployment requires more than predictive accuracy. Models should integrate with with EHRs and hospital information systems, provide interpretable outputs consistent with clinical reasoning, and function reliably under real-time and noisy conditions~\cite{blezek2021ai, wells2025practical}. These challenges, along with regulatory compliance, data privacy, and usability concerns, remain underexplored, leaving a translational gap between research prototypes and clinical adoption. Nevertheless, ECG-FMs hold considerable promise for integration into practice. For example, they could assist clinicians by generating preliminary diagnostic reports, prioritizing high-risk cases for review, or combining ECG-derived representations with multimodal patient data to inform treatment decisions~\cite{johnson2025artificial, martinez2023current}. Therefore, future research should emphasize prospective validation in clinical environments, development of user-centered interfaces, and collaboration with regulatory bodies to ensure safe and effective deployment. Addressing these challenges will be essential for ECG-FMs to evolve from promising research tools into practical decision-support systems capable of enhancing patient care.

\subsection*{Limited downstream tasks are tested.}

As illustrated in Figure 5, we categorize the applications of ECG-FMs into 11 types: (1) general cardiac arrhythmia classification/detection (CAC/CAD), (2) stress monitoring, (3) sleep monitoring (e.g., sleep stage and sleep apnea), (4) special cardiac diseases (e.g., LVSD, LVEF, ACTN, HCM), (5) patient information identification (e.g., age and gender), (6) zero-shot inference, (7) ECG synthesis, (8) blood parameter estimation (e.g., mPAP and BGL), (9) ECG question answering, (10) signal refinement (e.g., imputation and denoising), and (11) other tasks (e.g., emotion and arousal classification). Our analysis reveals that the predominant focus has been on classification tasks, with over 40\% of studies addressing CAC or CAD, while nearly 20\% target patient information identification and zero-shot classification. In contrast, fewer than 10\% of studies apply ECG-FMs to other cardiac diseases, raising concerns about the comprehensiveness of model evaluation and the breadth of clinical conditions being addressed. Notably, there remain many common but underexplored cardiac disease cohorts, such as hyperkalemia~\cite{galloway2019development} and hypertrophic cardiomyopathy~\cite{ko2020detection}, which warrant greater attention in ECG-FM development. Another limitation lies in the heavy reliance on a small set of public ECG datasets, such as PTB-XL, Chapman, CODE, CinC, and MIMIC. The scarcity of certain conditions in these datasets results in imbalanced categories, potentially biasing the detection of rare abnormalities. To address these limitations, future research should pursue a broader set of downstream tasks and leverage more diverse ECG datasets to fully unlock the potential of foundation models in advancing nuanced and clinically meaningful ECG analysis.

\begin{figure}[ht]
\centering
\includegraphics[width=.7\linewidth]{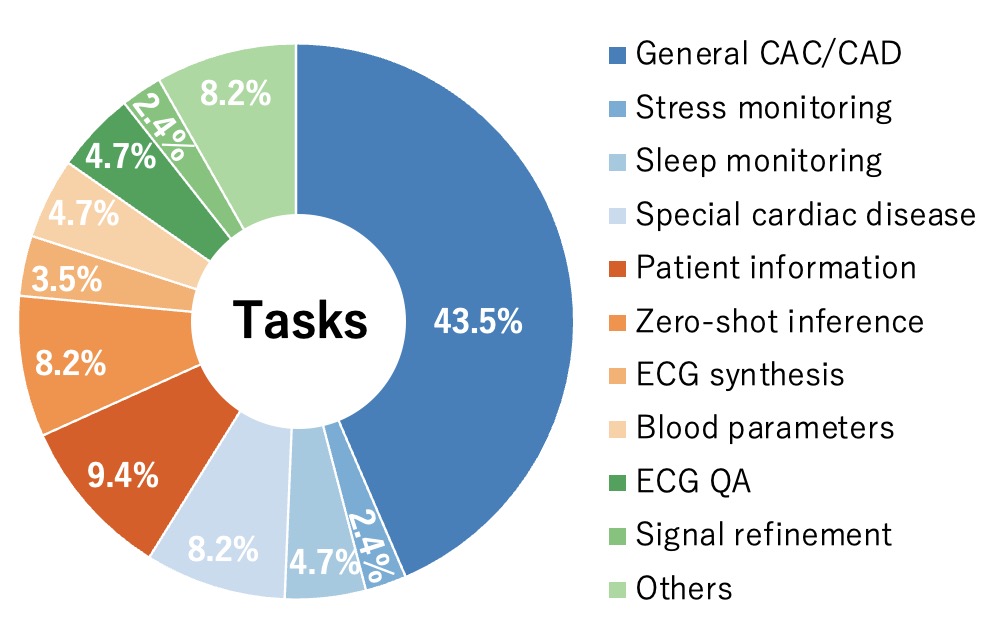}
\caption{\textbf{Types of downstream tasks in the reviewed papers.}}
\label{fig:pie_chart}
\end{figure}

\subsection*{Model weights are not shared.}

Out of 60 studies on ECG-FM, only 3 papers~\cite{yang2024biot, liu2023joint, mckeen2024ecg} have shared their saved weight files of the pre-trained model. These papers were able to do so because they utilized public datasets. Notably, the LLM community has a strong tradition of supporting and sharing their pre-trained FMs, as evidenced by the LLaMA-series, Baichuan-series, and Qwen-series models. However, the ECG research community lacks a clear guide on how to share pre-trained weights due to data privacy and ethics concerns, which hampers the development of ECG-FMs. Additionally, data isolation remains a significant challenge, with hospitals being hesitant to share their limited ECG data and collaborate to build a comprehensive ECG database. As a result, there is a lack of diversity in available ECG data for researchers and cardiologists. Furthermore, while high-tech companies typically develop and share state-of-the-art FMs, hospitals and medical institutions struggle to make full use of their real-time ECG data due to limited computing resources. Therefore, it is crucial to encourage collaboration among these institutions to drive the development of ECG-FMs and fully leverage valuable ECG data.

\subsection*{Data quality is not considered.}

LLMs typically employ stringent data quality filtering processes to ensure that they are trained on high-quality textual data. This approach enhances the models' ability to generate accurate and coherent outputs by learning correct syntax, semantics, and contextual nuances from clean and reliable sources. However, when this methodology is directly applied to the development of ECG-FMs, it presents significant challenges. In real clinical practice, a substantial portion of ECG data is of suboptimal quality due to various factors such as patient movement, poor electrode placement, electrical interference, and physiological artifacts like muscle contractions or respiratory movements~\cite{hannun2019cardiologist, liu2018open, alday2020classification}. If ECG-FMs are trained exclusively on high-quality, artifact-free ECG recordings, they may lack the robustness to handle the noisy and diverse data commonly encountered in clinical environments. This limitation can lead to significant degradation in model performance when deployed in real-world settings, potentially resulting in inaccurate interpretations or missed detections of critical cardiac abnormalities.

\subsection*{Benchmarking framework remains unstandardized.}

A critical limitation in the current development of ECG-FMs lies in the lack of standardized benchmarking frameworks. Although recent studies have demonstrated strong performance across diverse datasets and tasks, results are typically reported using heterogeneous datasets, evaluation metrics, and experimental protocols, which hinders fair comparison across models and complicates the assessment of generalizability~\cite{strodthoff2020deep}. This fragmentation not only obscures the relative strengths and weaknesses of different approaches but also raises concerns regarding robustness when models are applied to heterogeneous patient populations and real-world clinical settings. Establishing standardized evaluation suites, incorporating cross-dataset testing protocols, unified performance metrics, and task-specific benchmarks, would thus provide a more rigorous foundation for structuring future ECG-FM research. While no comprehensive benchmarking standard currently exists for ECG-FMs, collaborative efforts between the clinical and machine learning communities could play a pivotal role in curating representative datasets and defining shared evaluation criteria.

\subsection*{The larger the better? We don't know.}

The question of whether larger models inherently lead to better performance in the realm of ECG-FMs remains unanswered. Empirical studies have undeniably shown that LLMs adhere to scaling laws—principles indicating that model performance improves predictably with increases in parameters, training data size, and computational resources~\cite{hassid2024larger}. These scaling laws suggest that as models grow, they can capture more intricate patterns and nuances within the data, leading to enhanced performance across various tasks. However, the applicability of these scaling laws to the ECG domain is still under investigation. ECG data possess unique characteristics, such as complex temporal dynamics and significant inter-individual variability, which may influence how model size affects performance. Unlike text data, ECG signals require models to interpret subtle physiological variations that might not benefit simply from increased parameter counts. Therefore, it is uncertain whether larger ECG models will provide better diagnostic accuracy or generalization capabilities.

\subsection*{ECG-FMs may have the same problem as LLMs.}

LLMs are often difficult to interpret, which complicates efforts to elucidate the mechanisms underlying their representation learning and output generation~\cite{zhang2023siren}. Despite ongoing theoretical efforts, a comprehensive understanding of their pre-training dynamics and generalization abilities remains limited. AS a result, LLMs are prone to producing factually incorrect yet coherent and plausible outputs, a phenomenon commonly known as hallucination~\cite{farquhar2024detecting}. Similar concerns arise for ECG-FMs, where the generation of misleading information poses serious risks in clinical contexts. In ECG analysis, such hallucinations may manifest as unreliable diagnostic cues, potentially leading to misinterpretation and adverse patient outcomes. This problem is compounded by the black-box nature of ECG-FMs, which provide limited transparency into their decision-making processes and hinder both clinical adoption and regulatory approval. To address these limitations, explainable AI (XAI) techniques have emerged as promising solutions. For example, Grad-CAM~\cite{jahmunah2022explainable} can highlight salient temporal or spatial features in ECG signals, while SHAP~\cite{anand2022explainable} can attribute predictive importance to specific waveform segments. By offering visual or quantitative explanations aligned with clinical reasoning, these methods improve interpretability and foster trust. Although their use in ECG analysis remains at an early stage, integrating XAI frameworks into ECG-FMs could play a pivotal role in mitigating hallucination risks, enhancing transparency, and enabling safer deployment in real-world healthcare environments.

\subsection*{When and how are ECG-FMs ready for cardiologists?}

While ECG-FMs have demonstrated impressive versatility and adaptability for various cardiac-related tasks, they are still in the experimental phase. Ongoing research is focused on enhancing their performance and interpretability for clinical use. Notably, existing ECG AI tools like CardioAI and IBM Watson~\cite{kaul2020history} have already gained regulatory approvals and are being implemented into clinical practice with user-friendly interfaces. However, there is still progress to be made before ECG-FMs become widely accessible and seamlessly integrated into real-time workflows. On the other hand, how can ECG-FMs really help cardiologists? First, ECG-FMs can greatly aid in cardiology report generation, potentially improving the precision of reports. Furthermore, these advanced systems have the potential to support the detection of challenging or rare cardiac conditions, offering valuable insights and suggestions in cases where cardiologists may struggle to make a confident diagnosis.

\section*{Conclusions}

Recent advancements in FMs have been revolutionizing the field of ECG analysis. In this survey, we provide a practical guide to FMs specifically tailored for various ECG-based diagnostic scenarios from a methodology-centric perspective. We categorize ECG-FMs based on key elements such as model architecture, pre-training paradigm, and adaptation technique. Additionally, we summarize the ECG-related datasets and FM benchmarks, while addressing current challenges and outlining future directions. We believe that this survey will facilitate and inspire further innovation in the application of FMs to ECG analysis.

\section*{Author Contributions}
Yu Han conducted the literature collection, formal analysis, and drafted the manuscript. Vittorio Murino and Xiaofeng Liu contributed to methodology development, and critical manuscript revision. Xiang Zhang assisted with visualization, validation, and critical manuscript revision. Cheng Ding conceived and designed the study, provided supervision throughout all stages, contributed to methodological refinement, critically revised the manuscript. All authors reviewed and approved the final version of the manuscript.

\section*{Acknowledgment}
None.

\section*{Funding}
This research received no external funding.

\section*{Conflict of Interest}
The authors declare that they have no conflicts of interest relevant to this work.
\bibliography{main}

\begin{thebibliography}{100}
\urlstyle{rm}
\expandafter\ifx\csname url\endcsname\relax
  \def\url#1{\texttt{#1}}\fi
\expandafter\ifx\csname urlprefix\endcsname\relax\def\urlprefix{URL }\fi
\expandafter\ifx\csname doiprefix\endcsname\relax\def\doiprefix{DOI: }\fi
\providecommand{\bibinfo}[2]{#2}
\providecommand{\eprint}[2][]{\url{#2}}

\bibitem{kirchenbauer2023watermark}
\bibinfo{author}{Kirchenbauer, J.} \emph{et~al.}
\newblock \bibinfo{title}{A watermark for large language models}.
\newblock In \emph{\bibinfo{booktitle}{International Conference on Machine Learning}}, \bibinfo{pages}{17061--17084} (\bibinfo{organization}{PMLR}, \bibinfo{year}{2023}).

\bibitem{morin2021artificial}
\bibinfo{author}{Morin, O.} \emph{et~al.}
\newblock \bibinfo{journal}{\bibinfo{title}{An artificial intelligence framework integrating longitudinal electronic health records with real-world data enables continuous pan-cancer prognostication}}.
\newblock {\emph{\JournalTitle{Nature Cancer}}} \textbf{\bibinfo{volume}{2}}, \bibinfo{pages}{709--722} (\bibinfo{year}{2021}).

\bibitem{vemulapalliknowledge}
\bibinfo{author}{Vemulapalli, R.} \emph{et~al.}
\newblock \bibinfo{title}{Knowledge transfer from vision foundation models for efficient training of small task-specific models}.
\newblock In \emph{\bibinfo{booktitle}{Forty-first International Conference on Machine Learning}}.

\bibitem{fei2022towards}
\bibinfo{author}{Fei, N.} \emph{et~al.}
\newblock \bibinfo{journal}{\bibinfo{title}{Towards artificial general intelligence via a multimodal foundation model}}.
\newblock {\emph{\JournalTitle{Nature Communications}}} \textbf{\bibinfo{volume}{13}}, \bibinfo{pages}{3094} (\bibinfo{year}{2022}).

\bibitem{xu2023mplug}
\bibinfo{author}{Xu, H.} \emph{et~al.}
\newblock \bibinfo{title}{mplug-2: A modularized multi-modal foundation model across text, image and video}.
\newblock In \emph{\bibinfo{booktitle}{International Conference on Machine Learning}}, \bibinfo{pages}{38728--38748} (\bibinfo{organization}{PMLR}, \bibinfo{year}{2023}).

\bibitem{achiam2023gpt}
\bibinfo{author}{Achiam, J.} \emph{et~al.}
\newblock \bibinfo{journal}{\bibinfo{title}{Gpt-4 technical report}}.
\newblock {\emph{\JournalTitle{arXiv preprint arXiv:2303.08774}}}  (\bibinfo{year}{2023}).

\bibitem{touvron2023llama}
\bibinfo{author}{Touvron, H.} \emph{et~al.}
\newblock \bibinfo{journal}{\bibinfo{title}{Llama: Open and efficient foundation language models}}.
\newblock {\emph{\JournalTitle{arXiv preprint arXiv:2302.13971}}}  (\bibinfo{year}{2023}).

\bibitem{dubey2024llama}
\bibinfo{author}{Dubey, A.} \emph{et~al.}
\newblock \bibinfo{journal}{\bibinfo{title}{The llama 3 herd of models}}.
\newblock {\emph{\JournalTitle{arXiv preprint arXiv:2407.21783}}}  (\bibinfo{year}{2024}).

\bibitem{guo2024multi}
\bibinfo{author}{Guo, L.~L.} \emph{et~al.}
\newblock \bibinfo{journal}{\bibinfo{title}{A multi-center study on the adaptability of a shared foundation model for electronic health records}}.
\newblock {\emph{\JournalTitle{npj Digital Medicine}}} \textbf{\bibinfo{volume}{7}}, \bibinfo{pages}{171} (\bibinfo{year}{2024}).

\bibitem{melnyk2023reprogramming}
\bibinfo{author}{Melnyk, I.} \emph{et~al.}
\newblock \bibinfo{title}{Reprogramming pretrained language models for antibody sequence infilling}.
\newblock In \emph{\bibinfo{booktitle}{International Conference on Machine Learning}}, \bibinfo{pages}{24398--24419} (\bibinfo{organization}{PMLR}, \bibinfo{year}{2023}).

\bibitem{moor2023foundation}
\bibinfo{author}{Moor, M.} \emph{et~al.}
\newblock \bibinfo{journal}{\bibinfo{title}{Foundation models for generalist medical artificial intelligence}}.
\newblock {\emph{\JournalTitle{Nature}}} \textbf{\bibinfo{volume}{616}}, \bibinfo{pages}{259--265} (\bibinfo{year}{2023}).

\bibitem{noseworthy2022artificial}
\bibinfo{author}{Noseworthy, P.~A.} \emph{et~al.}
\newblock \bibinfo{journal}{\bibinfo{title}{Artificial intelligence-guided screening for atrial fibrillation using electrocardiogram during sinus rhythm: a prospective non-randomised interventional trial}}.
\newblock {\emph{\JournalTitle{The Lancet}}} \textbf{\bibinfo{volume}{400}}, \bibinfo{pages}{1206--1212} (\bibinfo{year}{2022}).

\bibitem{scholte2024scoping}
\bibinfo{author}{Scholte, N.~T.} \emph{et~al.}
\newblock \bibinfo{journal}{\bibinfo{title}{A scoping review on advancements in noninvasive wearable technology for heart failure management}}.
\newblock {\emph{\JournalTitle{npj Digital Medicine}}} \textbf{\bibinfo{volume}{7}}, \bibinfo{pages}{1--15} (\bibinfo{year}{2024}).

\bibitem{liu2024learning}
\bibinfo{author}{Liu, W.} \emph{et~al.}
\newblock \bibinfo{journal}{\bibinfo{title}{Learning representations for multi-lead electrocardiograms from morphology-rhythm contrast}}.
\newblock {\emph{\JournalTitle{IEEE Transactions on Instrumentation and Measurement}}}  (\bibinfo{year}{2024}).

\bibitem{georgieva2024examination}
\bibinfo{author}{Georgieva-Tsaneva, G.}, \bibinfo{author}{Gospodinova, E.} \& \bibinfo{author}{Cheshmedzhiev, K.}
\newblock \bibinfo{journal}{\bibinfo{title}{Examination of cardiac activity with ecg monitoring using heart rate variability methods}}.
\newblock {\emph{\JournalTitle{Diagnostics}}} \textbf{\bibinfo{volume}{14}}, \bibinfo{pages}{926} (\bibinfo{year}{2024}).

\bibitem{raghu2022contrastive}
\bibinfo{author}{Raghu, A.}, \bibinfo{author}{Chandak, P.}, \bibinfo{author}{Alam, R.}, \bibinfo{author}{Guttag, J.} \& \bibinfo{author}{Stultz, C.}
\newblock \bibinfo{title}{Contrastive pre-training for multimodal medical time series}.
\newblock In \emph{\bibinfo{booktitle}{NeurIPS 2022 Workshop on Learning from Time Series for Health}} (\bibinfo{year}{2022}).

\bibitem{wornow2023shaky}
\bibinfo{author}{Wornow, M.} \emph{et~al.}
\newblock \bibinfo{journal}{\bibinfo{title}{The shaky foundations of large language models and foundation models for electronic health records}}.
\newblock {\emph{\JournalTitle{npj Digital Medicine}}} \textbf{\bibinfo{volume}{6}}, \bibinfo{pages}{135} (\bibinfo{year}{2023}).

\bibitem{qi2022cybertwin}
\bibinfo{author}{Qi, W.} \& \bibinfo{author}{Su, H.}
\newblock \bibinfo{journal}{\bibinfo{title}{A cybertwin based multimodal network for ecg patterns monitoring using deep learning}}.
\newblock {\emph{\JournalTitle{IEEE Transactions on Industrial Informatics}}} \textbf{\bibinfo{volume}{18}}, \bibinfo{pages}{6663--6670} (\bibinfo{year}{2022}).

\bibitem{lan2023performer}
\bibinfo{author}{Lan, E.}
\newblock \bibinfo{title}{Performer: A novel ppg-to-ecg reconstruction transformer for a digital biomarker of cardiovascular disease detection}.
\newblock In \emph{\bibinfo{booktitle}{Proceedings of the IEEE/CVF Winter Conference on Applications of Computer Vision}}, \bibinfo{pages}{1991--1999} (\bibinfo{year}{2023}).

\bibitem{chung2023text}
\bibinfo{author}{Chung, H.} \emph{et~al.}
\newblock \bibinfo{title}{Text-to-ecg: 12-lead electrocardiogram synthesis conditioned on clinical text reports}.
\newblock In \emph{\bibinfo{booktitle}{ICASSP 2023-2023 IEEE International Conference on Acoustics, Speech and Signal Processing (ICASSP)}}, \bibinfo{pages}{1--5} (\bibinfo{organization}{IEEE}, \bibinfo{year}{2023}).

\bibitem{acosta2022multimodal}
\bibinfo{author}{Acosta, J.~N.}, \bibinfo{author}{Falcone, G.~J.}, \bibinfo{author}{Rajpurkar, P.} \& \bibinfo{author}{Topol, E.~J.}
\newblock \bibinfo{journal}{\bibinfo{title}{Multimodal biomedical ai}}.
\newblock {\emph{\JournalTitle{Nature Medicine}}} \textbf{\bibinfo{volume}{28}}, \bibinfo{pages}{1773--1784} (\bibinfo{year}{2022}).

\bibitem{awais2025foundation}
\bibinfo{author}{Awais, M.} \emph{et~al.}
\newblock \bibinfo{journal}{\bibinfo{title}{Foundation models defining a new era in vision: a survey and outlook}}.
\newblock {\emph{\JournalTitle{IEEE Transactions on Pattern Analysis and Machine Intelligence}}}  (\bibinfo{year}{2025}).

\bibitem{he2024foundation}
\bibinfo{author}{He, Y.} \emph{et~al.}
\newblock \bibinfo{journal}{\bibinfo{title}{Foundation model for advancing healthcare: Challenges, opportunities and future directions}}.
\newblock {\emph{\JournalTitle{IEEE Reviews in Biomedical Engineering}}}  (\bibinfo{year}{2024}).

\bibitem{khan2025comprehensive}
\bibinfo{author}{Khan, W.} \emph{et~al.}
\newblock \bibinfo{journal}{\bibinfo{title}{A comprehensive survey of foundation models in medicine}}.
\newblock {\emph{\JournalTitle{IEEE Reviews in Biomedical Engineering}}}  (\bibinfo{year}{2025}).

\bibitem{liang2024foundation}
\bibinfo{author}{Liang, Y.} \emph{et~al.}
\newblock \bibinfo{title}{Foundation models for time series analysis: A tutorial and survey}.
\newblock In \emph{\bibinfo{booktitle}{Proceedings of the 30th ACM SIGKDD Conference on Knowledge Discovery and Data Mining}}, \bibinfo{pages}{6555--6565} (\bibinfo{year}{2024}).

\bibitem{page2021prisma}
\bibinfo{author}{Page, M.~J.} \emph{et~al.}
\newblock \bibinfo{journal}{\bibinfo{title}{The prisma 2020 statement: an updated guideline for reporting systematic reviews}}.
\newblock {\emph{\JournalTitle{bmj}}} \textbf{\bibinfo{volume}{372}} (\bibinfo{year}{2021}).

\bibitem{bommasani2021opportunities}
\bibinfo{author}{Bommasani, R.} \emph{et~al.}
\newblock \bibinfo{journal}{\bibinfo{title}{On the opportunities and risks of foundation models}}.
\newblock {\emph{\JournalTitle{arXiv preprint arXiv:2108.07258}}}  (\bibinfo{year}{2021}).

\bibitem{wei2022emergent}
\bibinfo{author}{Wei, J.} \emph{et~al.}
\newblock \bibinfo{journal}{\bibinfo{title}{Emergent abilities of large language models}}.
\newblock {\emph{\JournalTitle{arXiv preprint arXiv:2206.07682}}}  (\bibinfo{year}{2022}).

\bibitem{bai2025qwen2}
\bibinfo{author}{Bai, S.} \emph{et~al.}
\newblock \bibinfo{journal}{\bibinfo{title}{Qwen2. 5-vl technical report}}.
\newblock {\emph{\JournalTitle{arXiv preprint arXiv:2502.13923}}}  (\bibinfo{year}{2025}).

\bibitem{liu2024deepseek}
\bibinfo{author}{Liu, A.} \emph{et~al.}
\newblock \bibinfo{journal}{\bibinfo{title}{Deepseek-v3 technical report}}.
\newblock {\emph{\JournalTitle{arXiv preprint arXiv:2412.19437}}}  (\bibinfo{year}{2024}).

\bibitem{devlin2018bert}
\bibinfo{author}{Devlin, J.}
\newblock \bibinfo{journal}{\bibinfo{title}{Bert: Pre-training of deep bidirectional transformers for language understanding}}.
\newblock {\emph{\JournalTitle{arXiv preprint arXiv:1810.04805}}}  (\bibinfo{year}{2018}).

\bibitem{raffel2020exploring}
\bibinfo{author}{Raffel, C.} \emph{et~al.}
\newblock \bibinfo{journal}{\bibinfo{title}{Exploring the limits of transfer learning with a unified text-to-text transformer}}.
\newblock {\emph{\JournalTitle{Journal of machine learning research}}} \textbf{\bibinfo{volume}{21}}, \bibinfo{pages}{1--67} (\bibinfo{year}{2020}).

\bibitem{liu2019roberta}
\bibinfo{author}{Liu, Y.}
\newblock \bibinfo{journal}{\bibinfo{title}{Roberta: A robustly optimized bert pretraining approach}}.
\newblock {\emph{\JournalTitle{arXiv preprint arXiv:1907.11692}}}  (\bibinfo{year}{2019}).

\bibitem{sanh2019distilbert}
\bibinfo{author}{Sanh, V.}
\newblock \bibinfo{journal}{\bibinfo{title}{Distilbert, a distilled version of bert: Smaller, faster, cheaper and lighter}}.
\newblock {\emph{\JournalTitle{arXiv preprint arXiv:1910.01108}}}  (\bibinfo{year}{2019}).

\bibitem{clark2020electra}
\bibinfo{author}{Clark, K.}
\newblock \bibinfo{journal}{\bibinfo{title}{Electra: Pre-training text encoders as discriminators rather than generators}}.
\newblock {\emph{\JournalTitle{arXiv preprint arXiv:2003.10555}}}  (\bibinfo{year}{2020}).

\bibitem{simonyan2014very}
\bibinfo{author}{Simonyan, K.} \& \bibinfo{author}{Zisserman, A.}
\newblock \bibinfo{journal}{\bibinfo{title}{Very deep convolutional networks for large-scale image recognition}}.
\newblock {\emph{\JournalTitle{arXiv preprint arXiv:1409.1556}}}  (\bibinfo{year}{2014}).

\bibitem{tan2019efficientnet}
\bibinfo{author}{Tan, M.}
\newblock \bibinfo{journal}{\bibinfo{title}{Efficientnet: Rethinking model scaling for convolutional neural networks}}.
\newblock {\emph{\JournalTitle{arXiv preprint arXiv:1905.11946}}}  (\bibinfo{year}{2019}).

\bibitem{dosovitskiy2020image}
\bibinfo{author}{Dosovitskiy, A.}
\newblock \bibinfo{journal}{\bibinfo{title}{An image is worth 16x16 words: Transformers for image recognition at scale}}.
\newblock {\emph{\JournalTitle{arXiv preprint arXiv:2010.11929}}}  (\bibinfo{year}{2020}).

\bibitem{liu2021swin}
\bibinfo{author}{Liu, Z.} \emph{et~al.}
\newblock \bibinfo{title}{Swin transformer: Hierarchical vision transformer using shifted windows}.
\newblock In \emph{\bibinfo{booktitle}{Proceedings of the IEEE/CVF international conference on computer vision}}, \bibinfo{pages}{10012--10022} (\bibinfo{year}{2021}).

\bibitem{wang2020linformer}
\bibinfo{author}{Wang, S.}, \bibinfo{author}{Li, B.~Z.}, \bibinfo{author}{Khabsa, M.}, \bibinfo{author}{Fang, H.} \& \bibinfo{author}{Ma, H.}
\newblock \bibinfo{journal}{\bibinfo{title}{Linformer: Self-attention with linear complexity}}.
\newblock {\emph{\JournalTitle{arXiv preprint arXiv:2006.04768}}}  (\bibinfo{year}{2020}).

\bibitem{touvron2021training}
\bibinfo{author}{Touvron, H.} \emph{et~al.}
\newblock \bibinfo{title}{Training data-efficient image transformers \& distillation through attention}.
\newblock In \emph{\bibinfo{booktitle}{International conference on machine learning}}, \bibinfo{pages}{10347--10357} (\bibinfo{organization}{PMLR}, \bibinfo{year}{2021}).

\bibitem{he2022masked}
\bibinfo{author}{He, K.} \emph{et~al.}
\newblock \bibinfo{title}{Masked autoencoders are scalable vision learners}.
\newblock In \emph{\bibinfo{booktitle}{Proceedings of the IEEE/CVF conference on computer vision and pattern recognition}}, \bibinfo{pages}{16000--16009} (\bibinfo{year}{2022}).

\bibitem{radford2021learning}
\bibinfo{author}{Radford, A.} \emph{et~al.}
\newblock \bibinfo{title}{Learning transferable visual models from natural language supervision}.
\newblock In \emph{\bibinfo{booktitle}{International conference on machine learning}}, \bibinfo{pages}{8748--8763} (\bibinfo{organization}{PMLR}, \bibinfo{year}{2021}).

\bibitem{kirillov2023segment}
\bibinfo{author}{Kirillov, A.} \emph{et~al.}
\newblock \bibinfo{title}{Segment anything}.
\newblock In \emph{\bibinfo{booktitle}{Proceedings of the IEEE/CVF International Conference on Computer Vision}}, \bibinfo{pages}{4015--4026} (\bibinfo{year}{2023}).

\bibitem{ma2024segment}
\bibinfo{author}{Ma, J.} \emph{et~al.}
\newblock \bibinfo{journal}{\bibinfo{title}{Segment anything in medical images}}.
\newblock {\emph{\JournalTitle{Nature Communications}}} \textbf{\bibinfo{volume}{15}}, \bibinfo{pages}{654} (\bibinfo{year}{2024}).

\bibitem{guzhov2022audioclip}
\bibinfo{author}{Guzhov, A.}, \bibinfo{author}{Raue, F.}, \bibinfo{author}{Hees, J.} \& \bibinfo{author}{Dengel, A.}
\newblock \bibinfo{title}{Audioclip: Extending clip to image, text and audio}.
\newblock In \emph{\bibinfo{booktitle}{ICASSP 2022-2022 IEEE International Conference on Acoustics, Speech and Signal Processing (ICASSP)}}, \bibinfo{pages}{976--980} (\bibinfo{organization}{IEEE}, \bibinfo{year}{2022}).

\bibitem{ghiasi2022scaling}
\bibinfo{author}{Ghiasi, G.}, \bibinfo{author}{Gu, X.}, \bibinfo{author}{Cui, Y.} \& \bibinfo{author}{Lin, T.-Y.}
\newblock \bibinfo{title}{Scaling open-vocabulary image segmentation with image-level labels}.
\newblock In \emph{\bibinfo{booktitle}{European Conference on Computer Vision}}, \bibinfo{pages}{540--557} (\bibinfo{organization}{Springer}, \bibinfo{year}{2022}).

\bibitem{jin2023large}
\bibinfo{author}{Jin, M.} \emph{et~al.}
\newblock \bibinfo{journal}{\bibinfo{title}{Large models for time series and spatio-temporal data: A survey and outlook}}.
\newblock {\emph{\JournalTitle{arXiv preprint arXiv:2310.10196}}}  (\bibinfo{year}{2023}).

\bibitem{yang2021voice2series}
\bibinfo{author}{Yang, C.-H.~H.}, \bibinfo{author}{Tsai, Y.-Y.} \& \bibinfo{author}{Chen, P.-Y.}
\newblock \bibinfo{title}{Voice2series: Reprogramming acoustic models for time series classification}.
\newblock In \emph{\bibinfo{booktitle}{International conference on machine learning}}, \bibinfo{pages}{11808--11819} (\bibinfo{organization}{PMLR}, \bibinfo{year}{2021}).

\bibitem{yue2022ts2vec}
\bibinfo{author}{Yue, Z.} \emph{et~al.}
\newblock \bibinfo{title}{Ts2vec: Towards universal representation of time series}.
\newblock In \emph{\bibinfo{booktitle}{Proceedings of the AAAI Conference on Artificial Intelligence}}, vol.~\bibinfo{volume}{36}, \bibinfo{pages}{8980--8987} (\bibinfo{year}{2022}).

\bibitem{dong2024simmtm}
\bibinfo{author}{Dong, J.} \emph{et~al.}
\newblock \bibinfo{journal}{\bibinfo{title}{Simmtm: A simple pre-training framework for masked time-series modeling}}.
\newblock {\emph{\JournalTitle{Advances in Neural Information Processing Systems}}} \textbf{\bibinfo{volume}{36}} (\bibinfo{year}{2024}).

\bibitem{liu2024unitime}
\bibinfo{author}{Liu, X.} \emph{et~al.}
\newblock \bibinfo{title}{Unitime: A language-empowered unified model for cross-domain time series forecasting}.
\newblock In \emph{\bibinfo{booktitle}{Proceedings of the ACM on Web Conference 2024}}, \bibinfo{pages}{4095--4106} (\bibinfo{year}{2024}).

\bibitem{gao2024units}
\bibinfo{author}{Gao, S.} \emph{et~al.}
\newblock \bibinfo{journal}{\bibinfo{title}{Units: Building a unified time series model}}.
\newblock {\emph{\JournalTitle{arXiv preprint arXiv:2403.00131}}}  (\bibinfo{year}{2024}).

\bibitem{das2023decoder}
\bibinfo{author}{Das, A.}, \bibinfo{author}{Kong, W.}, \bibinfo{author}{Sen, R.} \& \bibinfo{author}{Zhou, Y.}
\newblock \bibinfo{journal}{\bibinfo{title}{A decoder-only foundation model for time-series forecasting}}.
\newblock {\emph{\JournalTitle{arXiv preprint arXiv:2310.10688}}}  (\bibinfo{year}{2023}).

\bibitem{garza2023timegpt}
\bibinfo{author}{Garza, A.} \& \bibinfo{author}{Mergenthaler-Canseco, M.}
\newblock \bibinfo{journal}{\bibinfo{title}{Timegpt-1}}.
\newblock {\emph{\JournalTitle{arXiv preprint arXiv:2310.03589}}}  (\bibinfo{year}{2023}).

\bibitem{ekambaram2024ttms}
\bibinfo{author}{Ekambaram, V.} \emph{et~al.}
\newblock \bibinfo{journal}{\bibinfo{title}{Ttms: Fast multi-level tiny time mixers for improved zero-shot and few-shot forecasting of multivariate time series}}.
\newblock {\emph{\JournalTitle{arXiv preprint arXiv:2401.03955}}}  (\bibinfo{year}{2024}).

\bibitem{ekambaram2023tsmixer}
\bibinfo{author}{Ekambaram, V.}, \bibinfo{author}{Jati, A.}, \bibinfo{author}{Nguyen, N.}, \bibinfo{author}{Sinthong, P.} \& \bibinfo{author}{Kalagnanam, J.}
\newblock \bibinfo{title}{Tsmixer: Lightweight mlp-mixer model for multivariate time series forecasting}.
\newblock In \emph{\bibinfo{booktitle}{Proceedings of the 29th ACM SIGKDD Conference on Knowledge Discovery and Data Mining}}, \bibinfo{pages}{459--469} (\bibinfo{year}{2023}).

\bibitem{liu2022contrastive}
\bibinfo{author}{Liu, X.} \emph{et~al.}
\newblock \bibinfo{title}{When do contrastive learning signals help spatio-temporal graph forecasting?}
\newblock In \emph{\bibinfo{booktitle}{Proceedings of the 30th international conference on advances in geographic information systems}}, \bibinfo{pages}{1--12} (\bibinfo{year}{2022}).

\bibitem{li2022mining}
\bibinfo{author}{Li, R.} \emph{et~al.}
\newblock \bibinfo{title}{Mining spatio-temporal relations via self-paced graph contrastive learning}.
\newblock In \emph{\bibinfo{booktitle}{Proceedings of the 28th ACM SIGKDD Conference on Knowledge Discovery and Data Mining}}, \bibinfo{pages}{936--944} (\bibinfo{year}{2022}).

\bibitem{liu2024spatial}
\bibinfo{author}{Liu, C.} \emph{et~al.}
\newblock \bibinfo{journal}{\bibinfo{title}{Spatial-temporal large language model for traffic prediction}}.
\newblock {\emph{\JournalTitle{arXiv preprint arXiv:2401.10134}}}  (\bibinfo{year}{2024}).

\bibitem{huang2024repurposing}
\bibinfo{author}{Huang, N.}, \bibinfo{author}{Wang, H.}, \bibinfo{author}{He, Z.}, \bibinfo{author}{Zitnik, M.} \& \bibinfo{author}{Zhang, X.}
\newblock \bibinfo{journal}{\bibinfo{title}{Repurposing foundation model for generalizable medical time series classification}}.
\newblock {\emph{\JournalTitle{arXiv preprint arXiv:2410.03794}}}  (\bibinfo{year}{2024}).

\bibitem{wang2024contrast}
\bibinfo{author}{Wang, Y.}, \bibinfo{author}{Han, Y.}, \bibinfo{author}{Wang, H.} \& \bibinfo{author}{Zhang, X.}
\newblock \bibinfo{journal}{\bibinfo{title}{Contrast everything: A hierarchical contrastive framework for medical time-series}}.
\newblock {\emph{\JournalTitle{Advances in Neural Information Processing Systems}}} \textbf{\bibinfo{volume}{36}} (\bibinfo{year}{2024}).

\bibitem{vaswani2017attention}
\bibinfo{author}{Vaswani, A.}
\newblock \bibinfo{journal}{\bibinfo{title}{Attention is all you need}}.
\newblock {\emph{\JournalTitle{Advances in Neural Information Processing Systems}}}  (\bibinfo{year}{2017}).

\bibitem{kolodiazhnyi2024oneformer3d}
\bibinfo{author}{Kolodiazhnyi, M.}, \bibinfo{author}{Vorontsova, A.}, \bibinfo{author}{Konushin, A.} \& \bibinfo{author}{Rukhovich, D.}
\newblock \bibinfo{title}{Oneformer3d: One transformer for unified point cloud segmentation}.
\newblock In \emph{\bibinfo{booktitle}{Proceedings of the IEEE/CVF Conference on Computer Vision and Pattern Recognition}}, \bibinfo{pages}{20943--20953} (\bibinfo{year}{2024}).

\bibitem{lin2024fraudgt}
\bibinfo{author}{Lin, J.} \emph{et~al.}
\newblock \bibinfo{title}{Fraudgt: a simple, effective, and efficient graph transformer for financial fraud detection}.
\newblock In \emph{\bibinfo{booktitle}{Proceedings of the 5th ACM International Conference on AI in Finance}}, \bibinfo{pages}{292--300} (\bibinfo{year}{2024}).

\bibitem{peng2023study}
\bibinfo{author}{Peng, C.} \emph{et~al.}
\newblock \bibinfo{journal}{\bibinfo{title}{A study of generative large language model for medical research and healthcare}}.
\newblock {\emph{\JournalTitle{NPJ digital medicine}}} \textbf{\bibinfo{volume}{6}}, \bibinfo{pages}{210} (\bibinfo{year}{2023}).

\bibitem{khan2024comprehensive}
\bibinfo{author}{Khan, W.} \emph{et~al.}
\newblock \bibinfo{journal}{\bibinfo{title}{A comprehensive survey of foundation models in medicine}}.
\newblock {\emph{\JournalTitle{arXiv preprint arXiv:2406.10729}}}  (\bibinfo{year}{2024}).

\bibitem{yang2024biot}
\bibinfo{author}{Yang, C.}, \bibinfo{author}{Westover, M.} \& \bibinfo{author}{Sun, J.}
\newblock \bibinfo{journal}{\bibinfo{title}{Biot: Biosignal transformer for cross-data learning in the wild}}.
\newblock {\emph{\JournalTitle{Advances in Neural Information Processing Systems}}} \textbf{\bibinfo{volume}{36}} (\bibinfo{year}{2024}).

\bibitem{liu2023biosignal}
\bibinfo{author}{Liu, C.}, \bibinfo{author}{Ma, Y.}, \bibinfo{author}{Kothur, K.}, \bibinfo{author}{Nikpour, A.} \& \bibinfo{author}{Kavehei, O.}
\newblock \bibinfo{journal}{\bibinfo{title}{Biosignal copilot: Leveraging the power of llms in drafting reports for biomedical signals}}.
\newblock {\emph{\JournalTitle{medRxiv}}} \bibinfo{pages}{2023--06} (\bibinfo{year}{2023}).

\bibitem{yu2024ecg}
\bibinfo{author}{Yu, H.}, \bibinfo{author}{Guo, P.} \& \bibinfo{author}{Sano, A.}
\newblock \bibinfo{journal}{\bibinfo{title}{Ecg semantic integrator (esi): A foundation ecg model pretrained with llm-enhanced cardiological text}}.
\newblock {\emph{\JournalTitle{arXiv preprint arXiv:2405.19366}}}  (\bibinfo{year}{2024}).

\bibitem{liu2024zero}
\bibinfo{author}{Liu, C.} \emph{et~al.}
\newblock \bibinfo{journal}{\bibinfo{title}{Zero-shot ecg classification with multimodal learning and test-time clinical knowledge enhancement}}.
\newblock {\emph{\JournalTitle{arXiv preprint arXiv:2403.06659}}}  (\bibinfo{year}{2024}).

\bibitem{yu2023zero}
\bibinfo{author}{Yu, H.}, \bibinfo{author}{Guo, P.} \& \bibinfo{author}{Sano, A.}
\newblock \bibinfo{title}{Zero-shot ecg diagnosis with large language models and retrieval-augmented generation}.
\newblock In \emph{\bibinfo{booktitle}{Machine Learning for Health (ML4H)}}, \bibinfo{pages}{650--663} (\bibinfo{organization}{PMLR}, \bibinfo{year}{2023}).

\bibitem{khunte2024automated}
\bibinfo{author}{Khunte, A.} \emph{et~al.}
\newblock \bibinfo{journal}{\bibinfo{title}{Automated diagnostic reports from images of electrocardiograms at the point-of-care}}.
\newblock {\emph{\JournalTitle{medRxiv}}}  (\bibinfo{year}{2024}).

\bibitem{vaid2023foundational}
\bibinfo{author}{Vaid, A.} \emph{et~al.}
\newblock \bibinfo{journal}{\bibinfo{title}{A foundational vision transformer improves diagnostic performance for electrocardiograms}}.
\newblock {\emph{\JournalTitle{NPJ Digital Medicine}}} \textbf{\bibinfo{volume}{6}}, \bibinfo{pages}{108} (\bibinfo{year}{2023}).

\bibitem{gaudilliere2021generative}
\bibinfo{author}{Gaudilliere, P.~L.} \emph{et~al.}
\newblock \bibinfo{title}{Generative pre-trained transformer for cardiac abnormality detection}.
\newblock In \emph{\bibinfo{booktitle}{2021 Computing in Cardiology (CinC)}}, vol.~\bibinfo{volume}{48}, \bibinfo{pages}{1--4} (\bibinfo{organization}{IEEE}, \bibinfo{year}{2021}).

\bibitem{zhou2023masked}
\bibinfo{author}{Zhou, Y.} \emph{et~al.}
\newblock \bibinfo{journal}{\bibinfo{title}{Masked transformer for electrocardiogram classification}}.
\newblock {\emph{\JournalTitle{arXiv preprint arXiv:2309.07136}}}  (\bibinfo{year}{2023}).

\bibitem{na2024guiding}
\bibinfo{author}{Na, Y.}, \bibinfo{author}{Park, M.}, \bibinfo{author}{Tae, Y.} \& \bibinfo{author}{Joo, S.}
\newblock \bibinfo{journal}{\bibinfo{title}{Guiding masked representation learning to capture spatio-temporal relationship of electrocardiogram}}.
\newblock {\emph{\JournalTitle{arXiv preprint arXiv:2402.09450}}}  (\bibinfo{year}{2024}).

\bibitem{hu2023spatiotemporal}
\bibinfo{author}{Hu, R.}, \bibinfo{author}{Chen, J.} \& \bibinfo{author}{Zhou, L.}
\newblock \bibinfo{journal}{\bibinfo{title}{Spatiotemporal self-supervised representation learning from multi-lead ecg signals}}.
\newblock {\emph{\JournalTitle{Biomedical Signal Processing and Control}}} \textbf{\bibinfo{volume}{84}}, \bibinfo{pages}{104772} (\bibinfo{year}{2023}).

\bibitem{zhang2022maefe}
\bibinfo{author}{Zhang, H.} \emph{et~al.}
\newblock \bibinfo{journal}{\bibinfo{title}{Maefe: Masked autoencoders family of electrocardiogram for self-supervised pretraining and transfer learning}}.
\newblock {\emph{\JournalTitle{IEEE Transactions on Instrumentation and Measurement}}} \textbf{\bibinfo{volume}{72}}, \bibinfo{pages}{1--15} (\bibinfo{year}{2022}).

\bibitem{sawano2022masked}
\bibinfo{author}{Sawano, S.} \emph{et~al.}
\newblock \bibinfo{title}{Masked autoencoder-based self-supervised learning for electrocardiograms to detect left ventricular systolic dysfunction.}
\newblock In \emph{\bibinfo{booktitle}{NeurIPS 2022 Workshop on Learning from Time Series for Health}} (\bibinfo{year}{2022}).

\bibitem{yang2022masked}
\bibinfo{author}{Yang, S.}, \bibinfo{author}{Lian, C.} \& \bibinfo{author}{Zeng, Z.}
\newblock \bibinfo{title}{Masked autoencoder for ecg representation learning}.
\newblock In \emph{\bibinfo{booktitle}{2022 12th International Conference on Information Science and Technology (ICIST)}}, \bibinfo{pages}{95--98} (\bibinfo{organization}{IEEE}, \bibinfo{year}{2022}).

\bibitem{abbaspourazad2023large}
\bibinfo{author}{Abbaspourazad, S.} \emph{et~al.}
\newblock \bibinfo{journal}{\bibinfo{title}{Large-scale training of foundation models for wearable biosignals}}.
\newblock {\emph{\JournalTitle{arXiv preprint arXiv:2312.05409}}}  (\bibinfo{year}{2023}).

\bibitem{bao2021beit}
\bibinfo{author}{Bao, H.}, \bibinfo{author}{Dong, L.}, \bibinfo{author}{Piao, S.} \& \bibinfo{author}{Wei, F.}
\newblock \bibinfo{journal}{\bibinfo{title}{Beit: Bert pre-training of image transformers}}.
\newblock {\emph{\JournalTitle{arXiv preprint arXiv:2106.08254}}}  (\bibinfo{year}{2021}).

\bibitem{lee2020biobert}
\bibinfo{author}{Lee, J.} \emph{et~al.}
\newblock \bibinfo{journal}{\bibinfo{title}{Biobert: a pre-trained biomedical language representation model for biomedical text mining}}.
\newblock {\emph{\JournalTitle{Bioinformatics}}} \textbf{\bibinfo{volume}{36}}, \bibinfo{pages}{1234--1240} (\bibinfo{year}{2020}).

\bibitem{alsentzer2019publicly}
\bibinfo{author}{Alsentzer, E.} \emph{et~al.}
\newblock \bibinfo{journal}{\bibinfo{title}{Publicly available clinical bert embeddings}}.
\newblock {\emph{\JournalTitle{arXiv preprint arXiv:1904.03323}}}  (\bibinfo{year}{2019}).

\bibitem{yasunaga2022linkbert}
\bibinfo{author}{Yasunaga, M.}, \bibinfo{author}{Leskovec, J.} \& \bibinfo{author}{Liang, P.}
\newblock \bibinfo{journal}{\bibinfo{title}{Linkbert: Pretraining language models with document links}}.
\newblock {\emph{\JournalTitle{arXiv preprint arXiv:2203.15827}}}  (\bibinfo{year}{2022}).

\bibitem{deng2009imagenet}
\bibinfo{author}{Deng, J.} \emph{et~al.}
\newblock \bibinfo{title}{Imagenet: A large-scale hierarchical image database}.
\newblock In \emph{\bibinfo{booktitle}{2009 IEEE conference on computer vision and pattern recognition}}, \bibinfo{pages}{248--255} (\bibinfo{organization}{Ieee}, \bibinfo{year}{2009}).

\bibitem{liu2024etp}
\bibinfo{author}{Liu, C.}, \bibinfo{author}{Wan, Z.}, \bibinfo{author}{Cheng, S.}, \bibinfo{author}{Zhang, M.} \& \bibinfo{author}{Arcucci, R.}
\newblock \bibinfo{title}{Etp: Learning transferable ecg representations via ecg-text pre-training}.
\newblock In \emph{\bibinfo{booktitle}{ICASSP 2024-2024 IEEE International Conference on Acoustics, Speech and Signal Processing (ICASSP)}}, \bibinfo{pages}{8230--8234} (\bibinfo{organization}{IEEE}, \bibinfo{year}{2024}).

\bibitem{li2024frozen}
\bibinfo{author}{Li, J.}, \bibinfo{author}{Liu, C.}, \bibinfo{author}{Cheng, S.}, \bibinfo{author}{Arcucci, R.} \& \bibinfo{author}{Hong, S.}
\newblock \bibinfo{title}{Frozen language model helps ecg zero-shot learning}.
\newblock In \emph{\bibinfo{booktitle}{Medical Imaging with Deep Learning}}, \bibinfo{pages}{402--415} (\bibinfo{organization}{PMLR}, \bibinfo{year}{2024}).

\bibitem{choi2023ecgbert}
\bibinfo{author}{Choi, S.} \emph{et~al.}
\newblock \bibinfo{journal}{\bibinfo{title}{Ecgbert: Understanding hidden language of ecgs with self-supervised representation learning}}.
\newblock {\emph{\JournalTitle{arXiv preprint arXiv:2306.06340}}}  (\bibinfo{year}{2023}).

\bibitem{radford2019language}
\bibinfo{author}{Radford, A.} \emph{et~al.}
\newblock \bibinfo{journal}{\bibinfo{title}{Language models are unsupervised multitask learners}}.
\newblock {\emph{\JournalTitle{OpenAI blog}}} \textbf{\bibinfo{volume}{1}}, \bibinfo{pages}{9} (\bibinfo{year}{2019}).

\bibitem{wang2022medclip}
\bibinfo{author}{Wang, Z.}, \bibinfo{author}{Wu, Z.}, \bibinfo{author}{Agarwal, D.} \& \bibinfo{author}{Sun, J.}
\newblock \bibinfo{journal}{\bibinfo{title}{Medclip: Contrastive learning from unpaired medical images and text}}.
\newblock {\emph{\JournalTitle{arXiv preprint arXiv:2210.10163}}}  (\bibinfo{year}{2022}).

\bibitem{zhang2023biomedclip}
\bibinfo{author}{Zhang, S.} \emph{et~al.}
\newblock \bibinfo{journal}{\bibinfo{title}{Biomedclip: a multimodal biomedical foundation model pretrained from fifteen million scientific image-text pairs}}.
\newblock {\emph{\JournalTitle{arXiv preprint arXiv:2303.00915}}}  (\bibinfo{year}{2023}).

\bibitem{liu2025multimodal}
\bibinfo{author}{Liu, Y.} \emph{et~al.}
\newblock \bibinfo{journal}{\bibinfo{title}{Multimodal medical image binding via shared text embeddings}}.
\newblock {\emph{\JournalTitle{arXiv preprint arXiv:2506.18072}}}  (\bibinfo{year}{2025}).

\bibitem{takahashi2025application}
\bibinfo{author}{Takahashi, J.} \emph{et~al.}
\newblock \bibinfo{journal}{\bibinfo{title}{Application of contrastive learning on ecg data: Evaluating performance in japanese and classification with around 100 labels}}.
\newblock {\emph{\JournalTitle{arXiv preprint arXiv:2504.09302}}}  (\bibinfo{year}{2025}).

\bibitem{yang2025qwen3}
\bibinfo{author}{Yang, A.} \emph{et~al.}
\newblock \bibinfo{journal}{\bibinfo{title}{Qwen3 technical report}}.
\newblock {\emph{\JournalTitle{arXiv preprint arXiv:2505.09388}}}  (\bibinfo{year}{2025}).

\bibitem{radford2018improving}
\bibinfo{author}{Radford, A.}, \bibinfo{author}{Narasimhan, K.}, \bibinfo{author}{Salimans, T.}, \bibinfo{author}{Sutskever, I.} \emph{et~al.}
\newblock \bibinfo{journal}{\bibinfo{title}{Improving language understanding by generative pre-training}}.
\newblock {\emph{\JournalTitle{San Francisco, CA, USA}}}  (\bibinfo{year}{2018}).

\bibitem{wang2025ecg}
\bibinfo{author}{Wang, X.} \emph{et~al.}
\newblock \bibinfo{journal}{\bibinfo{title}{Ecg-expert-qa: A benchmark for evaluating medical large language models in heart disease diagnosis}}.
\newblock {\emph{\JournalTitle{arXiv preprint arXiv:2502.17475}}}  (\bibinfo{year}{2025}).

\bibitem{yang2025ecg}
\bibinfo{author}{Yang, K.} \emph{et~al.}
\newblock \bibinfo{journal}{\bibinfo{title}{Ecg-lm: Understanding electrocardiogram with a large language model}}.
\newblock {\emph{\JournalTitle{Health Data Science}}} \textbf{\bibinfo{volume}{5}}, \bibinfo{pages}{0221} (\bibinfo{year}{2025}).

\bibitem{shazeer2020glu}
\bibinfo{author}{Shazeer, N.}
\newblock \bibinfo{journal}{\bibinfo{title}{Glu variants improve transformer}}.
\newblock {\emph{\JournalTitle{arXiv preprint arXiv:2002.05202}}}  (\bibinfo{year}{2020}).

\bibitem{su2024roformer}
\bibinfo{author}{Su, J.} \emph{et~al.}
\newblock \bibinfo{journal}{\bibinfo{title}{Roformer: Enhanced transformer with rotary position embedding}}.
\newblock {\emph{\JournalTitle{Neurocomputing}}} \textbf{\bibinfo{volume}{568}}, \bibinfo{pages}{127063} (\bibinfo{year}{2024}).

\bibitem{liu2024ecg}
\bibinfo{author}{Liu, L.}, \bibinfo{author}{Cui, G.}, \bibinfo{author}{Wan, C.}, \bibinfo{author}{Wu, D.} \& \bibinfo{author}{Li, Y.}
\newblock \bibinfo{title}{Ecg-llm: Leveraging large language models for low-quality ecg signal restoration}.
\newblock In \emph{\bibinfo{booktitle}{2024 IEEE International Conference on Bioinformatics and Biomedicine (BIBM)}}, \bibinfo{pages}{3537--3542} (\bibinfo{organization}{IEEE}, \bibinfo{year}{2024}).

\bibitem{liu2024large}
\bibinfo{author}{Liu, Z.} \emph{et~al.}
\newblock \bibinfo{title}{Large language models for cuffless blood pressure measurement from wearable biosignals}.
\newblock In \emph{\bibinfo{booktitle}{Proceedings of the 15th ACM International Conference on Bioinformatics, Computational Biology and Health Informatics}}, \bibinfo{pages}{1--11} (\bibinfo{year}{2024}).

\bibitem{ramesh2021zero}
\bibinfo{author}{Ramesh, A.} \emph{et~al.}
\newblock \bibinfo{title}{Zero-shot text-to-image generation}.
\newblock In \emph{\bibinfo{booktitle}{International conference on machine learning}}, \bibinfo{pages}{8821--8831} (\bibinfo{organization}{Pmlr}, \bibinfo{year}{2021}).

\bibitem{ramesh2022hierarchical}
\bibinfo{author}{Ramesh, A.}, \bibinfo{author}{Dhariwal, P.}, \bibinfo{author}{Nichol, A.}, \bibinfo{author}{Chu, C.} \& \bibinfo{author}{Chen, M.}
\newblock \bibinfo{journal}{\bibinfo{title}{Hierarchical text-conditional image generation with clip latents}}.
\newblock {\emph{\JournalTitle{arXiv preprint arXiv:2204.06125}}} \textbf{\bibinfo{volume}{1}}, \bibinfo{pages}{3} (\bibinfo{year}{2022}).

\bibitem{betker2023improvin}
\bibinfo{author}{Betker, J.} \emph{et~al.}
\newblock \bibinfo{journal}{\bibinfo{title}{Improving image generation with better captions}}.
\newblock {\emph{\JournalTitle{Computer Science. https://cdn. openai. com/papers/dall-e-3. pdf}}} \textbf{\bibinfo{volume}{2}}, \bibinfo{pages}{8} (\bibinfo{year}{2023}).

\bibitem{zhu2024can}
\bibinfo{author}{Zhu, L.}, \bibinfo{author}{Mou, W.}, \bibinfo{author}{Wu, K.}, \bibinfo{author}{Zhang, J.} \& \bibinfo{author}{Luo, P.}
\newblock \bibinfo{journal}{\bibinfo{title}{Can dall-e 3 reliably generate 12-lead ecgs and teaching illustrations?}}
\newblock {\emph{\JournalTitle{Cureus}}} \textbf{\bibinfo{volume}{16}} (\bibinfo{year}{2024}).

\bibitem{amri2023incorporating}
\bibinfo{author}{Amri, M.~M.} \& \bibinfo{author}{Hisan, U.~K.}
\newblock \bibinfo{journal}{\bibinfo{title}{Incorporating ai tools into medical education: harnessing the benefits of chatgpt and dall-e}}.
\newblock {\emph{\JournalTitle{Journal of Novel Engineering Science and Technology}}} \textbf{\bibinfo{volume}{2}}, \bibinfo{pages}{34--39} (\bibinfo{year}{2023}).

\bibitem{lalzary2023dual}
\bibinfo{author}{Lalzary, N.} \& \bibinfo{author}{Wolf, L.}
\newblock \bibinfo{title}{Dual contrastive learning for self-supervised ecg mapping to emotions and glucose levels}.
\newblock In \emph{\bibinfo{booktitle}{2023 IEEE SENSORS}}, \bibinfo{pages}{1--4} (\bibinfo{organization}{IEEE}, \bibinfo{year}{2023}).

\bibitem{brown2020language}
\bibinfo{author}{Brown, T.~B.}
\newblock \bibinfo{journal}{\bibinfo{title}{Language models are few-shot learners}}.
\newblock {\emph{\JournalTitle{arXiv preprint arXiv:2005.14165}}}  (\bibinfo{year}{2020}).

\bibitem{zhao2024ecg}
\bibinfo{author}{Zhao, Y.}, \bibinfo{author}{Kang, J.}, \bibinfo{author}{Zhang, T.}, \bibinfo{author}{Han, P.} \& \bibinfo{author}{Chen, T.}
\newblock \bibinfo{journal}{\bibinfo{title}{Ecg-chat: A large ecg-language model for cardiac disease diagnosis}}.
\newblock {\emph{\JournalTitle{arXiv preprint arXiv:2408.08849}}}  (\bibinfo{year}{2024}).

\bibitem{lan2025gem}
\bibinfo{author}{Lan, X.} \emph{et~al.}
\newblock \bibinfo{journal}{\bibinfo{title}{Gem: Empowering mllm for grounded ecg understanding with time series and images}}.
\newblock {\emph{\JournalTitle{arXiv preprint arXiv:2503.06073}}}  (\bibinfo{year}{2025}).

\bibitem{liu2024teach}
\bibinfo{author}{Liu, R.}, \bibinfo{author}{Bai, Y.}, \bibinfo{author}{Yue, X.} \& \bibinfo{author}{Zhang, P.}
\newblock \bibinfo{journal}{\bibinfo{title}{Teach multimodal llms to comprehend electrocardiographic images}}.
\newblock {\emph{\JournalTitle{arXiv preprint arXiv:2410.19008}}}  (\bibinfo{year}{2024}).

\bibitem{krizhevsky2012imagenet}
\bibinfo{author}{Krizhevsky, A.}, \bibinfo{author}{Sutskever, I.} \& \bibinfo{author}{Hinton, G.~E.}
\newblock \bibinfo{journal}{\bibinfo{title}{Imagenet classification with deep convolutional neural networks}}.
\newblock {\emph{\JournalTitle{Advances in neural information processing systems}}} \textbf{\bibinfo{volume}{25}} (\bibinfo{year}{2012}).

\bibitem{he2016deep}
\bibinfo{author}{He, K.}, \bibinfo{author}{Zhang, X.}, \bibinfo{author}{Ren, S.} \& \bibinfo{author}{Sun, J.}
\newblock \bibinfo{title}{Deep residual learning for image recognition}.
\newblock In \emph{\bibinfo{booktitle}{Proceedings of the IEEE conference on computer vision and pattern recognition}}, \bibinfo{pages}{770--778} (\bibinfo{year}{2016}).

\bibitem{liu2023joint}
\bibinfo{author}{Liu, W.} \emph{et~al.}
\newblock \bibinfo{journal}{\bibinfo{title}{A joint cross-dimensional contrastive learning framework for 12-lead ecgs and its heterogeneous deployment on soc}}.
\newblock {\emph{\JournalTitle{Computers in Biology and Medicine}}} \textbf{\bibinfo{volume}{152}}, \bibinfo{pages}{106390} (\bibinfo{year}{2023}).

\bibitem{gopal20213kg}
\bibinfo{author}{Gopal, B.} \emph{et~al.}
\newblock \bibinfo{title}{3kg: Contrastive learning of 12-lead electrocardiograms using physiologically-inspired augmentations}.
\newblock In \emph{\bibinfo{booktitle}{Machine Learning for Health}}, \bibinfo{pages}{156--167} (\bibinfo{organization}{PMLR}, \bibinfo{year}{2021}).

\bibitem{bo2022adversarial}
\bibinfo{author}{Bo, J.}, \bibinfo{author}{Huang, H.-W.}, \bibinfo{author}{Chan, A.} \& \bibinfo{author}{Traverso, G.}
\newblock \bibinfo{title}{Adversarial masking for pretraining ecg data improves downstream model generalizability}.
\newblock In \emph{\bibinfo{booktitle}{NeurIPS 2022 Workshop on Learning from Time Series for Health}}.

\bibitem{bo2022pretraining}
\bibinfo{author}{Bo, J.~Y.}, \bibinfo{author}{Huang, H.-W.}, \bibinfo{author}{Chan, A.} \& \bibinfo{author}{Traverso, G.}
\newblock \bibinfo{journal}{\bibinfo{title}{Pretraining ecg data with adversarial masking improves model generalizability for data-scarce tasks}}.
\newblock {\emph{\JournalTitle{arXiv preprint arXiv:2211.07889}}}  (\bibinfo{year}{2022}).

\bibitem{shi2024universal}
\bibinfo{author}{Shi, J.} \emph{et~al.}
\newblock \bibinfo{journal}{\bibinfo{title}{Universal 12-lead ecg representation for signal denoising and cardiovascular disease detection by fusing generative and contrastive learning}}.
\newblock {\emph{\JournalTitle{Biomedical Signal Processing and Control}}} \textbf{\bibinfo{volume}{94}}, \bibinfo{pages}{106253} (\bibinfo{year}{2024}).

\bibitem{phan2022multimodality}
\bibinfo{author}{Phan, T.} \emph{et~al.}
\newblock \bibinfo{title}{Multimodality multi-lead ecg arrhythmia classification using self-supervised learning}.
\newblock In \emph{\bibinfo{booktitle}{2022 IEEE-EMBS International Conference on Biomedical and Health Informatics (BHI)}}, \bibinfo{pages}{01--04} (\bibinfo{organization}{IEEE}, \bibinfo{year}{2022}).

\bibitem{mehari2022self}
\bibinfo{author}{Mehari, T.} \& \bibinfo{author}{Strodthoff, N.}
\newblock \bibinfo{journal}{\bibinfo{title}{Self-supervised representation learning from 12-lead ecg data}}.
\newblock {\emph{\JournalTitle{Computers in biology and medicine}}} \textbf{\bibinfo{volume}{141}}, \bibinfo{pages}{105114} (\bibinfo{year}{2022}).

\bibitem{grill2020bootstrap}
\bibinfo{author}{Grill, J.-B.} \emph{et~al.}
\newblock \bibinfo{journal}{\bibinfo{title}{Bootstrap your own latent-a new approach to self-supervised learning}}.
\newblock {\emph{\JournalTitle{Advances in neural information processing systems}}} \textbf{\bibinfo{volume}{33}}, \bibinfo{pages}{21271--21284} (\bibinfo{year}{2020}).

\bibitem{bardes2021vicreg}
\bibinfo{author}{Bardes, A.}, \bibinfo{author}{Ponce, J.} \& \bibinfo{author}{LeCun, Y.}
\newblock \bibinfo{journal}{\bibinfo{title}{Vicreg: Variance-invariance-covariance regularization for self-supervised learning}}.
\newblock {\emph{\JournalTitle{arXiv preprint arXiv:2105.04906}}}  (\bibinfo{year}{2021}).

\bibitem{le2023scl}
\bibinfo{author}{Le, D.}, \bibinfo{author}{Truong, S.}, \bibinfo{author}{Brijesh, P.}, \bibinfo{author}{Adjeroh, D.~A.} \& \bibinfo{author}{Le, N.}
\newblock \bibinfo{journal}{\bibinfo{title}{scl-st: Supervised contrastive learning with semantic transformations for multiple lead ecg arrhythmia classification}}.
\newblock {\emph{\JournalTitle{IEEE journal of biomedical and health informatics}}} \textbf{\bibinfo{volume}{27}}, \bibinfo{pages}{2818--2828} (\bibinfo{year}{2023}).

\bibitem{mehari2023towards}
\bibinfo{author}{Mehari, T.} \& \bibinfo{author}{Strodthoff, N.}
\newblock \bibinfo{journal}{\bibinfo{title}{Towards quantitative precision for ecg analysis: Leveraging state space models, self-supervision and patient metadata}}.
\newblock {\emph{\JournalTitle{IEEE Journal of Biomedical and Health Informatics}}}  (\bibinfo{year}{2023}).

\bibitem{lalam2023ecg}
\bibinfo{author}{Lalam, S.~K.} \emph{et~al.}
\newblock \bibinfo{journal}{\bibinfo{title}{Ecg representation learning with multi-modal ehr data}}.
\newblock {\emph{\JournalTitle{Transactions on Machine Learning Research}}}  (\bibinfo{year}{2023}).

\bibitem{yang2024masked}
\bibinfo{author}{Yang, S.} \emph{et~al.}
\newblock \bibinfo{journal}{\bibinfo{title}{Masked self-supervised ecg representation learning via multiview information bottleneck}}.
\newblock {\emph{\JournalTitle{Neural Computing and Applications}}} \textbf{\bibinfo{volume}{36}}, \bibinfo{pages}{7625--7637} (\bibinfo{year}{2024}).

\bibitem{mathew2024foundation}
\bibinfo{author}{Mathew, G.}, \bibinfo{author}{Barbosa, D.}, \bibinfo{author}{Prince, J.} \& \bibinfo{author}{Venkatraman, S.}
\newblock \bibinfo{journal}{\bibinfo{title}{Foundation models for cardiovascular disease detection via biosignals from digital stethoscopes}}.
\newblock {\emph{\JournalTitle{npj Cardiovascular Health}}} \textbf{\bibinfo{volume}{1}}, \bibinfo{pages}{25} (\bibinfo{year}{2024}).

\bibitem{shen2021partial}
\bibinfo{author}{Shen, Z.}, \bibinfo{author}{Liu, Z.}, \bibinfo{author}{Qin, J.}, \bibinfo{author}{Savvides, M.} \& \bibinfo{author}{Cheng, K.-T.}
\newblock \bibinfo{title}{Partial is better than all: Revisiting fine-tuning strategy for few-shot learning}.
\newblock In \emph{\bibinfo{booktitle}{Proceedings of the AAAI conference on artificial intelligence}}, vol.~\bibinfo{volume}{35}, \bibinfo{pages}{9594--9602} (\bibinfo{year}{2021}).

\bibitem{pittaras2017comparison}
\bibinfo{author}{Pittaras, N.}, \bibinfo{author}{Markatopoulou, F.}, \bibinfo{author}{Mezaris, V.} \& \bibinfo{author}{Patras, I.}
\newblock \bibinfo{title}{Comparison of fine-tuning and extension strategies for deep convolutional neural networks}.
\newblock In \emph{\bibinfo{booktitle}{MultiMedia Modeling: 23rd International Conference, MMM 2017, Reykjavik, Iceland, January 4-6, 2017, Proceedings, Part I 23}}, \bibinfo{pages}{102--114} (\bibinfo{organization}{Springer}, \bibinfo{year}{2017}).

\bibitem{hu2021lora}
\bibinfo{author}{Hu, E.~J.} \emph{et~al.}
\newblock \bibinfo{journal}{\bibinfo{title}{Lora: Low-rank adaptation of large language models}}.
\newblock {\emph{\JournalTitle{arXiv preprint arXiv:2106.09685}}}  (\bibinfo{year}{2021}).

\bibitem{suh2021learning}
\bibinfo{author}{Suh, J.} \emph{et~al.}
\newblock \bibinfo{title}{Learning ecg representations for multi-label classification of cardiac abnormalities}.
\newblock In \emph{\bibinfo{booktitle}{2021 Computing in Cardiology (CinC)}}, vol.~\bibinfo{volume}{48}, \bibinfo{pages}{1--4} (\bibinfo{organization}{IEEE}, \bibinfo{year}{2021}).

\bibitem{kiyasseh2021clocs}
\bibinfo{author}{Kiyasseh, D.}, \bibinfo{author}{Zhu, T.} \& \bibinfo{author}{Clifton, D.~A.}
\newblock \bibinfo{title}{Clocs: Contrastive learning of cardiac signals across space, time, and patients}.
\newblock In \emph{\bibinfo{booktitle}{International Conference on Machine Learning}}, \bibinfo{pages}{5606--5615} (\bibinfo{organization}{PMLR}, \bibinfo{year}{2021}).

\bibitem{rabbani2022contrastive}
\bibinfo{author}{Rabbani, S.} \& \bibinfo{author}{Khan, N.}
\newblock \bibinfo{journal}{\bibinfo{title}{Contrastive self-supervised learning for stress detection from ecg data}}.
\newblock {\emph{\JournalTitle{Bioengineering}}} \textbf{\bibinfo{volume}{9}}, \bibinfo{pages}{374} (\bibinfo{year}{2022}).

\bibitem{oh2022lead}
\bibinfo{author}{Oh, J.}, \bibinfo{author}{Chung, H.}, \bibinfo{author}{Kwon, J.-m.}, \bibinfo{author}{Hong, D.-g.} \& \bibinfo{author}{Choi, E.}
\newblock \bibinfo{title}{Lead-agnostic self-supervised learning for local and global representations of electrocardiogram}.
\newblock In \emph{\bibinfo{booktitle}{Conference on Health, Inference, and Learning}}, \bibinfo{pages}{338--353} (\bibinfo{organization}{PMLR}, \bibinfo{year}{2022}).

\bibitem{lan2022intra}
\bibinfo{author}{Lan, X.}, \bibinfo{author}{Ng, D.}, \bibinfo{author}{Hong, S.} \& \bibinfo{author}{Feng, M.}
\newblock \bibinfo{title}{Intra-inter subject self-supervised learning for multivariate cardiac signals}.
\newblock In \emph{\bibinfo{booktitle}{Proceedings of the AAAI Conference on Artificial Intelligence}}, vol.~\bibinfo{volume}{36}, \bibinfo{pages}{4532--4540} (\bibinfo{year}{2022}).

\bibitem{andersson2023augmentation}
\bibinfo{author}{Andersson, M.}, \bibinfo{author}{Nilsson, M.}, \bibinfo{author}{Flood, G.} \& \bibinfo{author}{{\AA}str{\"o}m, K.}
\newblock \bibinfo{title}{Augmentation strategies for self-supervised representation learning from electrocardiograms}.
\newblock In \emph{\bibinfo{booktitle}{2023 31st European Signal Processing Conference (EUSIPCO)}}, \bibinfo{pages}{1075--1079} (\bibinfo{organization}{IEEE}, \bibinfo{year}{2023}).

\bibitem{soltanieh2023distribution}
\bibinfo{author}{Soltanieh, S.}, \bibinfo{author}{Hashemi, J.} \& \bibinfo{author}{Etemad, A.}
\newblock \bibinfo{journal}{\bibinfo{title}{In-distribution and out-of-distribution self-supervised ecg representation learning for arrhythmia detection}}.
\newblock {\emph{\JournalTitle{IEEE Journal of Biomedical and Health Informatics}}}  (\bibinfo{year}{2023}).

\bibitem{liu2023dense}
\bibinfo{author}{Liu, W.} \emph{et~al.}
\newblock \bibinfo{journal}{\bibinfo{title}{Dense lead contrast for self-supervised representation learning of multilead electrocardiograms}}.
\newblock {\emph{\JournalTitle{Information Sciences}}} \textbf{\bibinfo{volume}{634}}, \bibinfo{pages}{189--205} (\bibinfo{year}{2023}).

\bibitem{lai2023practical}
\bibinfo{author}{Lai, J.} \emph{et~al.}
\newblock \bibinfo{journal}{\bibinfo{title}{Practical intelligent diagnostic algorithm for wearable 12-lead ecg via self-supervised learning on large-scale dataset}}.
\newblock {\emph{\JournalTitle{Nature Communications}}} \textbf{\bibinfo{volume}{14}}, \bibinfo{pages}{3741} (\bibinfo{year}{2023}).

\bibitem{nath2023towards}
\bibinfo{author}{Nath, R.~K.}, \bibinfo{author}{Tervonen, J.}, \bibinfo{author}{N{\"a}rv{\"a}inen, J.}, \bibinfo{author}{Pettersson, K.} \& \bibinfo{author}{M{\"a}ntyj{\"a}rvi, J.}
\newblock \bibinfo{title}{Towards self-supervised learning of ecg signal representation for the classification of acute stress types}.
\newblock In \emph{\bibinfo{booktitle}{Proceedings of the Great Lakes Symposium on VLSI 2023}}, \bibinfo{pages}{85--90} (\bibinfo{year}{2023}).

\bibitem{wang2023adversarial}
\bibinfo{author}{Wang, N.} \emph{et~al.}
\newblock \bibinfo{journal}{\bibinfo{title}{Adversarial spatiotemporal contrastive learning for electrocardiogram signals}}.
\newblock {\emph{\JournalTitle{IEEE Transactions on Neural Networks and Learning Systems}}}  (\bibinfo{year}{2023}).

\bibitem{thapa2024sleepfm}
\bibinfo{author}{Thapa, R.} \emph{et~al.}
\newblock \bibinfo{title}{Sleepfm: Multi-modal representation learning for sleep across ecg, eeg and respiratory signals}.
\newblock In \emph{\bibinfo{booktitle}{AAAI 2024 Spring Symposium on Clinical Foundation Models}} (\bibinfo{year}{2024}).

\bibitem{mckeen2024ecg}
\bibinfo{author}{McKeen, K.} \emph{et~al.}
\newblock \bibinfo{journal}{\bibinfo{title}{Ecg-fm: An open electrocardiogram foundation model}}.
\newblock {\emph{\JournalTitle{arXiv preprint arXiv:2408.05178}}}  (\bibinfo{year}{2024}).

\bibitem{li2024electrocardiogram}
\bibinfo{author}{Li, J.} \emph{et~al.}
\newblock \bibinfo{journal}{\bibinfo{title}{An electrocardiogram foundation model built on over 10 million recordings with external evaluation across multiple domains}}.
\newblock {\emph{\JournalTitle{arXiv preprint arXiv:2410.04133}}}  (\bibinfo{year}{2024}).

\bibitem{fu2024cardiogpt}
\bibinfo{author}{Fu, G.} \emph{et~al.}
\newblock \bibinfo{journal}{\bibinfo{title}{Cardiogpt: An ecg interpretation generation model}}.
\newblock {\emph{\JournalTitle{IEEE Access}}}  (\bibinfo{year}{2024}).

\bibitem{fang2024promoting}
\bibinfo{author}{Fang, C.} \emph{et~al.}
\newblock \bibinfo{journal}{\bibinfo{title}{Promoting cross-modal representations to improve multimodal foundation models for physiological signals}}.
\newblock {\emph{\JournalTitle{arXiv preprint arXiv:2410.16424}}}  (\bibinfo{year}{2024}).

\bibitem{blezek2021ai}
\bibinfo{author}{Blezek, D.~J.}, \bibinfo{author}{Olson-Williams, L.}, \bibinfo{author}{Missert, A.} \& \bibinfo{author}{Korfiatis, P.}
\newblock \bibinfo{journal}{\bibinfo{title}{Ai integration in the clinical workflow}}.
\newblock {\emph{\JournalTitle{Journal of Digital Imaging}}} \textbf{\bibinfo{volume}{34}}, \bibinfo{pages}{1435--1446} (\bibinfo{year}{2021}).

\bibitem{wells2025practical}
\bibinfo{author}{Wells, B.~J.} \emph{et~al.}
\newblock \bibinfo{journal}{\bibinfo{title}{A practical framework for appropriate implementation and review of artificial intelligence (fair-ai) in healthcare}}.
\newblock {\emph{\JournalTitle{NPJ digital medicine}}} \textbf{\bibinfo{volume}{8}}, \bibinfo{pages}{514} (\bibinfo{year}{2025}).

\bibitem{johnson2025artificial}
\bibinfo{author}{Johnson, L.} \emph{et~al.}
\newblock \bibinfo{journal}{\bibinfo{title}{Artificial intelligence for direct-to-physician reporting of ambulatory electrocardiography}}.
\newblock {\emph{\JournalTitle{Nature Medicine}}} \textbf{\bibinfo{volume}{31}}, \bibinfo{pages}{925--931} (\bibinfo{year}{2025}).

\bibitem{martinez2023current}
\bibinfo{author}{Mart{\'\i}nez-Sell{\'e}s, M.} \& \bibinfo{author}{Marina-Breysse, M.}
\newblock \bibinfo{journal}{\bibinfo{title}{Current and future use of artificial intelligence in electrocardiography}}.
\newblock {\emph{\JournalTitle{Journal of Cardiovascular Development and Disease}}} \textbf{\bibinfo{volume}{10}}, \bibinfo{pages}{175} (\bibinfo{year}{2023}).

\bibitem{galloway2019development}
\bibinfo{author}{Galloway, C.~D.} \emph{et~al.}
\newblock \bibinfo{journal}{\bibinfo{title}{Development and validation of a deep-learning model to screen for hyperkalemia from the electrocardiogram}}.
\newblock {\emph{\JournalTitle{JAMA cardiology}}} \textbf{\bibinfo{volume}{4}}, \bibinfo{pages}{428--436} (\bibinfo{year}{2019}).

\bibitem{ko2020detection}
\bibinfo{author}{Ko, W.-Y.} \emph{et~al.}
\newblock \bibinfo{journal}{\bibinfo{title}{Detection of hypertrophic cardiomyopathy using a convolutional neural network-enabled electrocardiogram}}.
\newblock {\emph{\JournalTitle{Journal of the American College of Cardiology}}} \textbf{\bibinfo{volume}{75}}, \bibinfo{pages}{722--733} (\bibinfo{year}{2020}).

\bibitem{hannun2019cardiologist}
\bibinfo{author}{Hannun, A.~Y.} \emph{et~al.}
\newblock \bibinfo{journal}{\bibinfo{title}{Cardiologist-level arrhythmia detection and classification in ambulatory electrocardiograms using a deep neural network}}.
\newblock {\emph{\JournalTitle{Nature medicine}}} \textbf{\bibinfo{volume}{25}}, \bibinfo{pages}{65--69} (\bibinfo{year}{2019}).

\bibitem{liu2018open}
\bibinfo{author}{Liu, F.} \emph{et~al.}
\newblock \bibinfo{journal}{\bibinfo{title}{An open access database for evaluating the algorithms of electrocardiogram rhythm and morphology abnormality detection}}.
\newblock {\emph{\JournalTitle{Journal of Medical Imaging and Health Informatics}}} \textbf{\bibinfo{volume}{8}}, \bibinfo{pages}{1368--1373} (\bibinfo{year}{2018}).

\bibitem{alday2020classification}
\bibinfo{author}{Alday, E. A.~P.} \emph{et~al.}
\newblock \bibinfo{journal}{\bibinfo{title}{Classification of 12-lead ecgs: the physionet/computing in cardiology challenge 2020}}.
\newblock {\emph{\JournalTitle{Physiological measurement}}} \textbf{\bibinfo{volume}{41}}, \bibinfo{pages}{124003} (\bibinfo{year}{2020}).

\bibitem{strodthoff2020deep}
\bibinfo{author}{Strodthoff, N.}, \bibinfo{author}{Wagner, P.}, \bibinfo{author}{Schaeffter, T.} \& \bibinfo{author}{Samek, W.}
\newblock \bibinfo{journal}{\bibinfo{title}{Deep learning for ecg analysis: Benchmarks and insights from ptb-xl}}.
\newblock {\emph{\JournalTitle{IEEE journal of biomedical and health informatics}}} \textbf{\bibinfo{volume}{25}}, \bibinfo{pages}{1519--1528} (\bibinfo{year}{2020}).

\bibitem{hassid2024larger}
\bibinfo{author}{Hassid, M.}, \bibinfo{author}{Remez, T.}, \bibinfo{author}{Gehring, J.}, \bibinfo{author}{Schwartz, R.} \& \bibinfo{author}{Adi, Y.}
\newblock \bibinfo{journal}{\bibinfo{title}{The larger the better? improved llm code-generation via budget reallocation}}.
\newblock {\emph{\JournalTitle{arXiv preprint arXiv:2404.00725}}}  (\bibinfo{year}{2024}).

\bibitem{zhang2023siren}
\bibinfo{author}{Zhang, Y.} \emph{et~al.}
\newblock \bibinfo{journal}{\bibinfo{title}{Siren's song in the ai ocean: a survey on hallucination in large language models}}.
\newblock {\emph{\JournalTitle{arXiv preprint arXiv:2309.01219}}}  (\bibinfo{year}{2023}).

\bibitem{farquhar2024detecting}
\bibinfo{author}{Farquhar, S.}, \bibinfo{author}{Kossen, J.}, \bibinfo{author}{Kuhn, L.} \& \bibinfo{author}{Gal, Y.}
\newblock \bibinfo{journal}{\bibinfo{title}{Detecting hallucinations in large language models using semantic entropy}}.
\newblock {\emph{\JournalTitle{Nature}}} \textbf{\bibinfo{volume}{630}}, \bibinfo{pages}{625--630} (\bibinfo{year}{2024}).

\bibitem{jahmunah2022explainable}
\bibinfo{author}{Jahmunah, V.}, \bibinfo{author}{Ng, E.~Y.}, \bibinfo{author}{Tan, R.-S.}, \bibinfo{author}{Oh, S.~L.} \& \bibinfo{author}{Acharya, U.~R.}
\newblock \bibinfo{journal}{\bibinfo{title}{Explainable detection of myocardial infarction using deep learning models with grad-cam technique on ecg signals}}.
\newblock {\emph{\JournalTitle{Computers in Biology and Medicine}}} \textbf{\bibinfo{volume}{146}}, \bibinfo{pages}{105550} (\bibinfo{year}{2022}).

\bibitem{anand2022explainable}
\bibinfo{author}{Anand, A.}, \bibinfo{author}{Kadian, T.}, \bibinfo{author}{Shetty, M.~K.} \& \bibinfo{author}{Gupta, A.}
\newblock \bibinfo{journal}{\bibinfo{title}{Explainable ai decision model for ecg data of cardiac disorders}}.
\newblock {\emph{\JournalTitle{Biomedical Signal Processing and Control}}} \textbf{\bibinfo{volume}{75}}, \bibinfo{pages}{103584} (\bibinfo{year}{2022}).

\bibitem{kaul2020history}
\bibinfo{author}{Kaul, V.}, \bibinfo{author}{Enslin, S.} \& \bibinfo{author}{Gross, S.~A.}
\newblock \bibinfo{journal}{\bibinfo{title}{History of artificial intelligence in medicine}}.
\newblock {\emph{\JournalTitle{Gastrointestinal endoscopy}}} \textbf{\bibinfo{volume}{92}}, \bibinfo{pages}{807--812} (\bibinfo{year}{2020}).

\end{thebibliography}

\end{document}